\documentclass[preprint,superscriptaddress,showpacs,preprintnumbers,amsmath,amssymb,aps,pre]{revtex4}
\usepackage{amsmath,amssymb}
\usepackage{epsfig}
\usepackage{graphics}
\usepackage{graphicx}
\usepackage{color}

\begin{document}

\title{Impact of Network Heterogeneity on Neuronal Synchronization}

\author{Javier Used}
\email[]{javier.used@urjc.es}
\affiliation{Nonlinear Dynamics, Chaos and Complex Systems Group, Departamento de
F\'{i}sica, Universidad Rey Juan Carlos, Tulip\'{a}n s/n, 28933 M\'{o}stoles, Madrid, Spain}

\author{Jes\'{u}s M. Seoane}
\affiliation{Nonlinear Dynamics, Chaos and Complex Systems Group, Departamento de
F\'{i}sica, Universidad Rey Juan Carlos, Tulip\'{a}n s/n, 28933 M\'{o}stoles, Madrid, Spain}

\author{Irina Bashkirtseva}
\affiliation{Institute of Natural Sciences and Mathematics, Ural Federal University, 620000, Lenina, 51, Ekaterinburg, Russia}

\author{Lev Ryashko}
\affiliation{Institute of Natural Sciences and Mathematics, Ural Federal University, 620000, Lenina, 51, Ekaterinburg, Russia}

\author{Miguel A.F. Sanju\'{a}n}
\affiliation{Nonlinear Dynamics, Chaos and Complex Systems Group, Departamento de
F\'{i}sica, Universidad Rey Juan Carlos, Tulip\'{a}n s/n, 28933 M\'{o}stoles, Madrid, Spain}

\date{\today}

\begin{abstract}
Synchronization dynamics is a phenomenon of great interest in many fields of science. One of the most important fields is neuron dynamics, as synchronization in certain regions of the brain is related to some of the most common mental illnesses. To study the impact of the network heterogeneity in the neuronal synchronization, we analyze a small-world network of non-identical Chialvo neurons that are electrically coupled. We introduce a mismatch in one of the model parameters to introduce the heterogeneity of the network. Our study examines the effects of this parameter mismatch, the noise intensity in the stochastic model, and the coupling strength between neurons on synchronization and firing frequency. We have identified critical values of noise intensity, parameter mismatch, and rewiring probability that facilitate effective synchronization within the network. Furthermore, we observe that the balance between excitatory and inhibitory connections plays a crucial role in achieving global synchronization. Our findings offer insights into the mechanisms driving synchronization dynamics in complex neuron networks.
\end{abstract}
\maketitle
\newpage
\section{Introduction} \label{introduction}

The phenomenon of collective synchronization occurs when a system of coupled oscillators spontaneously locks to a common behavior, independently of their individual characteristics. This phenomenon has been observed in many different scientific fields such as biology \cite{Peskin75,Liu97,aldridge99}, engineering \cite{Kourtchatov95,York91}, and ecology \cite{Buck76,Walker69}. In addition to all these examples, the synchronization of brain neurons is acquiring special interest in neuroscience because it is a fundamental neural mechanism \cite{Engel01,Uhlhaas06}. Different types of synchronous behaviors between neurons have been described, including complete synchronization, phase synchronization, and lag synchronization \cite{Boccaletti2018}. These behaviors have been related to various brain functions. For instance, the transition between in-phase and anti-phase synchronization is associated with locomotor brain rhythms \cite{Jia18,Song12}, while phase synchronization is related to neural integration and working memory \cite{Bambi00,Batista07,Varela01}. On the other hand, pathological cases of synchronization may result in serious brain disorders, such as essential tremor, Parkinson's disease, and epilepsy \cite{Pare90}. Beyond synchronization in the brain network, the firing frequency of neurons in the brain is of particular interest because it relates to neural coding. Recent studies have shown that the timing of spikes plays a critical role in encoding, representing, and processing knowledge and events in the brain \cite{Fujii96,MacLeod96}.
With all these considerations in mind, various models have been proposed to understand the synchronization of coupled neurons. Among them are continuous neuron models based on ordinary differential equations, such as Hodgkin-Huxley \cite{Hodgkin1952,Hodgkin1952_2}, FitzHugh-Nagumo \cite{FitzHugh1961}, Morris-Lecar \cite{morris:lecar}, and Hindmarsh-Rose \cite{Hindmarsh1982,Shilnikov2008}, which provide a realistic depiction of the electrophysiological processes within neurons. Conversely, discrete neuron models like Rulkov, Izhikevich, and Courbage-Nekorkin-Vdovin \cite{Rulkov01,Ibarz2011,Gir2013,KT96,Courbage2007} offer computational ease due to their simplicity, thereby reducing the computational workload.
In our work, we propose using the stochastic version of the Chialvo map \cite{Chialvo95} to understand synchronization in a neuron network. The effect of noise in excitable systems has been extensively studied for both white and colored noise \cite{zambrano:2010,LinGar04}, and it has been found that noise intensity is related to changes in the particular behavior of neuron models \cite{bruss:2023,BNR18,BNR_CSF18}. Moreover, it is also known that factors such as coupling strength \cite{Sun15,Shen08}, coupling forms \cite{Wang2011,Sriram2023}, and diversity \cite{Tang11} play key roles in the global synchronization of the network.  The choice of the stochastic Chialvo model over other models is motivated by the observation that critical values of noise and coupling significantly enhance network synchronization—a phenomenon not observed in other models in the same way as in this study. Some of those features have been studied in recent works. In Ref.~\cite{anjana24} the authors studied the dynamics of the system in a simple network configuration with identical neurons, while in Ref.~\cite{Kuznetsov24} the authors described the dynamics of the system when two or three non-identical neurons are connected. In this work, we take into account the heterogeneity of the network by introducing a parameter mismatch in the neuron model and consider different types of coupling, either excitatory or inhibitory but in a much more complicated network structure. The global synchronization of the network and the firing frequency will be studied using numerical methods for various characteristics of the neurons and the network itself. We have found critical values of noise intensity that enhance the synchronization of the neuron network. We observed that the presence of inhibitory couplings desynchronizes the system, while an increase in the rewiring probability improves the stability of the neuron network. All these features are explained in detail throughout the manuscript.
The organization of this paper is as follows. In Section~\ref{modeldescription1N}, we describe the stochastic Chialvo model. In Section~\ref{Nneuronsection}, we analyze the synchronization and the firing frequency in the small-world neuron network as a function of the mismatch between the neurons in one of the parameters of the model, the noise intensity an the proportion between the excitatory and the inhibitory couplings.  The main conclusions of our results are provided in Section~\ref{conclusions}.

\section{The stochastic Chialvo neuron model}
 \label{modeldescription1N}
To construct the neuron network, we will use the stochastic version of the map-based Chialvo model \cite{Used24}. This model is based in the deterministic Chialvo model \cite{Chialvo95} where the variable $x$ is related to an instantaneous membrane potential of the neuron, while the variable $y$ stands for the recovery current. In this case a random disturbance $\varepsilon\xi_t$ is added in the parameter $I$ of the acting ion current injected into the neuron. The parameter $\varepsilon$ represents the intensity of the noise, while $\xi_t$ is  uncorrelated white Gaussian noise with parameters $\langle\xi_t\rangle=0,\;\langle\xi^2_t\rangle=1$. The parameter $a$ is related to the time of recovery ($a < 1$), the activation-dependence of the recovery process is defined by the parameter $b$  ($b < 1$) with the offset value $c$.  All this information is summarized in Eq.~\ref{model1}:

\begin{equation}\label{model1}
\begin{array}{l}
  x_{t+1}=x_t^2 \exp(y_t-x_t)+I+\varepsilon\xi_t,\medskip\\
  y_{t+1}=a y_t-b x_t +c.
\end{array}
\end{equation}

\begin{figure}[ht!]
\centering
 \includegraphics[width=0.9\textwidth]{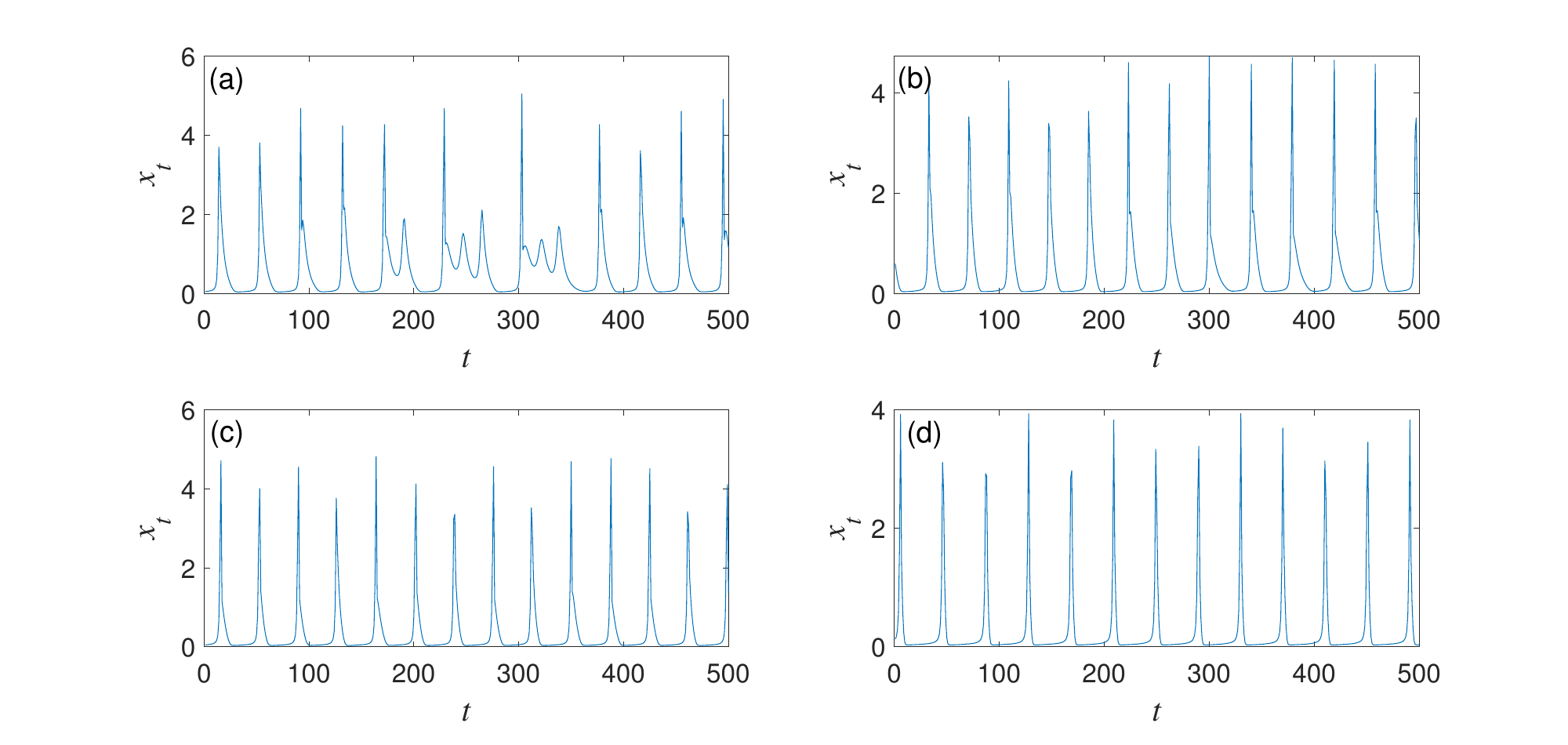}
 \caption{\textbf{Time series for the stochastic Chialvo neuron model}. Action potential $x_t$ is represented for different values of the parameter $b$: (a) $b$ = 0.18, (b) $b$ = 0.20, (c) $b$ = 0.22, (d) $b$ = 0.35. The rest of the parameters of the model remain fixed, $a=0.89$, $c=0.28$, $I=0.03$ and $\varepsilon = 1.5\cdot 10^{-3}$.}
\label{fig_figura_subplot_a_089_b_noise}
\end{figure}

This neuron model is capable of reproducing some of the basic features behind the firing dynamics of the neurons \cite{Chialvo95, Hoppensteadt86}. For example, by modifying the parameter $b$, one could obtain different dynamical behaviors: fixed point, periodic, quasi-periodic or chaotic dynamics. These basic dynamics of the neuron can be appreciated in Fig.~\ref{fig_figura_subplot_a_089_b_noise}(a-d) where the potential action, $x_t$, is represented for some typical values of the model parameters ($a$, $c$ and $I$) and for different values of the parameter $b$.

\begin{figure}[ht!]
\centering
\includegraphics[width=0.7\textwidth]{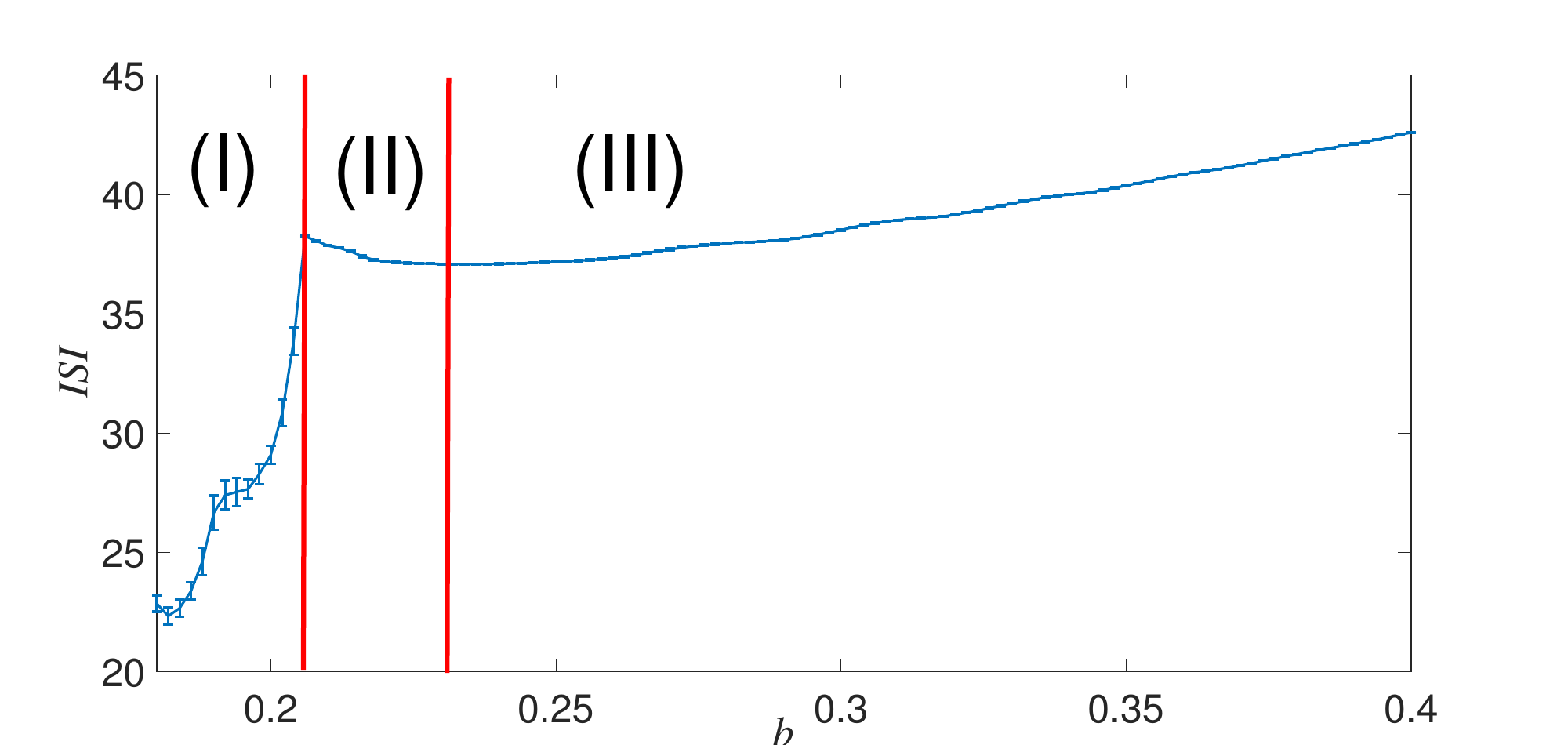}
 \caption{\textbf{Inter-spike interval ($ISI$) as a function of the parameter $b$}.  The values of the parameters are: $a=0.89$, $c=0.28$, $I=0.03,$ and the noise intensity is $\varepsilon = 1.5\cdot 10^{-3}$. The three regions where the relation between the parameters $b$ and $ISI$ is different are clearly appreciated and denoted in the plot.}
\label{fig_figura_ISI_b_a_89_prueba}
\end{figure}

To understand the influence of different parameters in the firing frequency of the neurons, we define the inter-spike interval, $ISI$, as the time between two consecutive spikes in the neuron time series. The dependence of the $ISI$ as a function of the parameter $b$ can be appreciated in Fig.~\ref{fig_figura_ISI_b_a_89_prueba}, where we represent the mean value of the $ISI$ as a function of $b$ and the error bars represent the standard deviation of the $ISI$. Every value of the plot is obtained as the mean value of $50$ simulations, each one starting from random conditions. It can be appreciated clearly three different regions: Region I: $0.18 \le b\le 0.206$, Region II: $0.206 \le b \le 0.23$, and Region III: $0.23 \le b$. In Region I, the value of $ISI$ increases as the parameter $b$ increases. In this region, small changes in the parameter $b$ imply big changes in the value of $ISI$. It can also be observed that the standard deviation of the $ISI$ is quite significant, which means that the $ISI$ is not very stable. In Region II, $b$ and $ISI$ have an inverse relationship, that is, when $b$ increases the mean value of $ISI$ decreases. Finally, in Region III the relationship between $b$ and $ISI$ is again direct and an increase in $b$ involves an increase in $ISI$. In the last two regions, it can be observed that the standard deviation of the parameter $ISI$ is negligible, being zero in many cases, that is the firing frequency of the neuron is almost constant. This means that, in spite of the differences in the amplitude in the signal that the model reproduces, the dynamical system is very stable in the frequency domain.

\begin{figure}[ht!]
\centering
\includegraphics[width=1\textwidth]{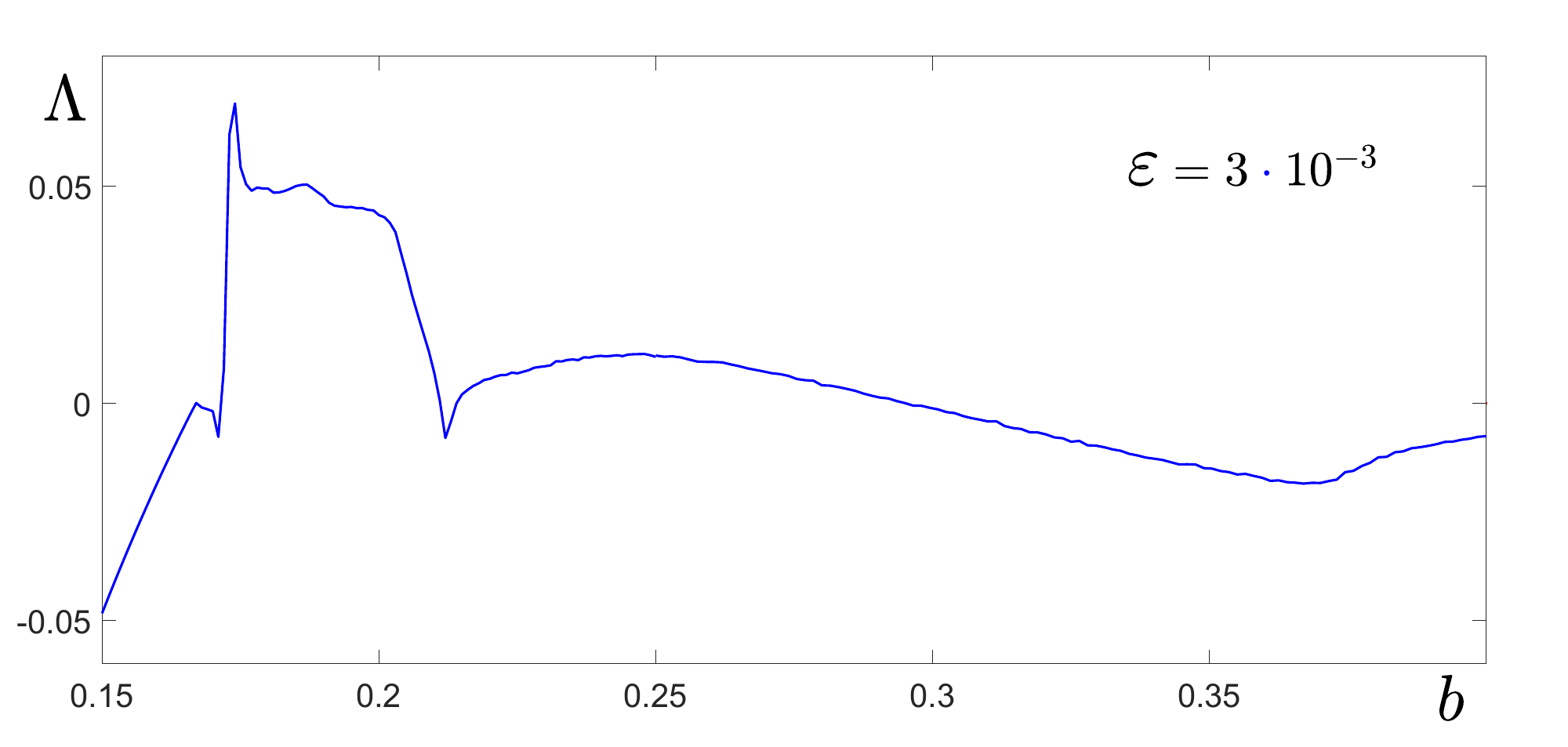}
\caption{\textbf{Largest Lyapunov exponent for the Chialvo neuron model versus the parameter $b$ for $\varepsilon=0.003$}. The rest of the parameters of the model are fixed to $a=0.89,\;I=0.03,$ and $\;c=0.28$.}
\label{Largest_Lyapunov}
\end{figure}

The diversity of these regimes can be also analyzed in Fig.~\ref{Largest_Lyapunov} where the largest Lyapunov exponent $\Lambda(b)$ is plotted for the case $\varepsilon=0.003$. Here, we have fixed $a=0.89,\;I=0.03,\;c=0.28$ and consider $b$ as the bifurcation parameter. As can be seen, a variation of $b$ implies crucial changes in the system (Eq.~\ref{model1}) dynamics with order-chaos transitions. Comparing Fig.~\ref{Largest_Lyapunov} with Fig.~\ref{fig_figura_ISI_b_a_89_prueba}, it can be appreciated how the region (I) corresponds to the highest values of $\Lambda$, that is when the system has a chaotic behavior. Regions (II) and (III) corresponds to low positive and negative values of $\Lambda$.

\section{Small-world neuron network}\label{Nneuronsection}
In a previous work \cite{Used24}, the effect of noise intensity and the mismatch in the parameter $b$ on synchronization and the firing frequency was anlyzed in a system constructed by connecting only two non-identical neurons. In this simple model we also considered two different types of couplings between the neurons: excitatory and inhibitory. To generalize the results obtained in this previous work we propose a similar study but in a small-world neuron network. With this purpose in mind, the first step is to construct the neuron network, that is, to connect mathematically a group of neurons. In our case, we connect the neurons in the well-known small-world network \cite{Watts98}. To build a small-world network of $N$ neurons, the first step is to order the $N$ neurons, and then connect each neuron with its $l$ first neighbors in a cyclical way. In this sense, the first node should be connected with the $l$ latest nodes. Finally, each node has a probability $p$ to rewire one of these edges with any other node. Thus, this kind of graph is characterized by three parameters: (i) $N$, the number of nodes; (ii) $l$, the number of first neighbors connected with each neuron; and (iii) $p$, the probability of rewiring an edge. This last parameter indicates the random nature of the graph, i.e., as $p$ increases the graph will be more random \cite{Watts98}. After the random rewiring of the network connections, each neuron will have a total number of connections $N_i$ that can change from one neuron to another. This parameter, $N_i$, plays an important role in the network, because the connection strength between the neuron $i$ and the neuron $j$ will be done by ($k/N_i$).

The randomness of the graph implies that, for the same choice of parameters, there might be different degrees of synchronization that could depend on the specific structure of the graph. This phenomenon makes sense in the real brain, since all brains have the same macroscopic structure but from a microscopic point of view the connection patterns are different for every person. For this reason, all throughout the paper, we will repeat the calculations $50$ times for every set of parameters of the network and we will represent the mean value of the parameter we are interested in.

\begin{figure}[ht!]
\centering
 \includegraphics[width=0.5\textwidth]{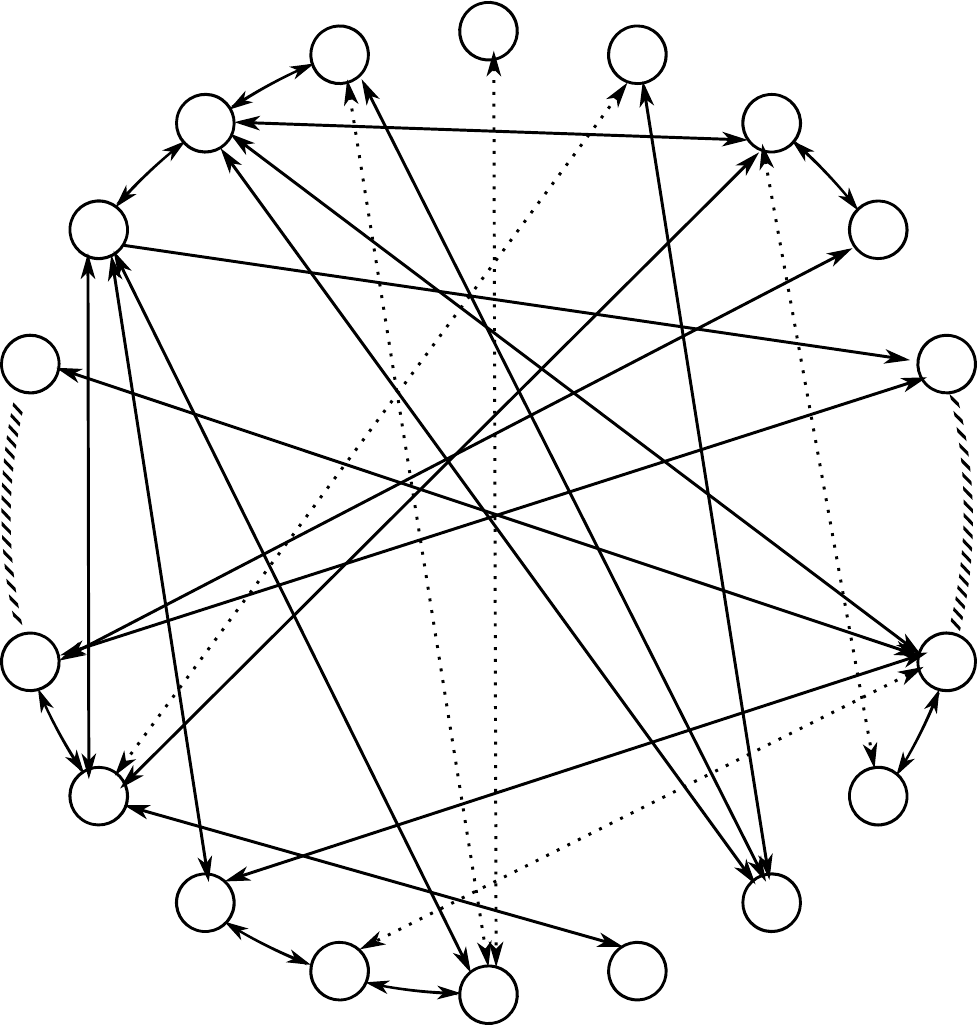}
 \caption{\textbf{Sketch of the neuron network}. In order to clarify the sketch we consider that initially each neuron is connected to its two closest neighbors ($l=1$). Solid arrows mark the excitatory couplings while the dashed arrows mark the inhibitory ones.} 
 \label{network}
\end{figure}

As previously mentioned, the connections between nodes in the graph represent the electrical couplings between the neurons. Mathematically,  the coupling is modeled by the addition of a term to our equations that represents the voltage difference between the two connected neurons. Besides, we have to consider if the coupling between every pair of neurons is excitatory or inhibitory. With all this information in mind, we can construct the mathematical representation of all the couplings for the neuron $i$:
\begin{equation}\label{eq_electricalcoupling}
	\dfrac{k}{N_i}\sum_j\delta_{ij}(x_{j,t-1}-x_{i,t-1}),
\end{equation}
where $i$ is the evaluated neuron, $j$ the remaining neurons, $k$ is the coupling strength and $\delta_{ij}$ is a coupling constant between neurons $i$ and $j$. If the connection between the neuron $i$ and the neuron $j$ is excitatory the parameter $\delta_{ij}=1$. Conversely, if the coupling is inhibitory $\delta_{ij}=-1$. This coupling constant would be equal to zero if neuron $i$ and neuron $j$ are not connected.

Summarizing all this previous information, see Fig.~\ref{network}, we proceed to build the complete model of the network. In this case, we will fix the same value for the coupling strength for all neurons, so that the model can be written as:
\begin{equation}\label{eq_network}
\begin{array}{l}
  x_{i,t+1}=x_{i,t}^2 \exp(y_{i,t}-x_{i,t})+I +\dfrac{k}{N_i}\sum_j\delta_{ij}(x_{j,t-1}-x_{i,t-1})+\varepsilon\xi_{i,t}\medskip\\
  y_{i,t+1}=a y_{i,t}-b_i x_{i,t} +c.\medskip\\
\end{array}.
\end{equation}

Once we have developed the mathematical model for the neuron network (Eq.~\ref{eq_network}), we proceed to the numerical simulations. In our case, we will set the mean value of the parameter $b$ at $b=0.35$  and will introduce a random mismatch in the parameter $b$ of some neurons. The maximum value of this mismatch would be ($\pm 0.01 b$). We introduce some small differences between the neurons using this value, while still considering them similar. It would be senseless to consider neurons located in the same brain region and part of the same local network as very different.  Introducing a mismatch in the parameter $b$ for every neuron and the fact that some of the links between the neurons are inhibitory connections introduce some instability in the neuron network, significantly affecting  the synchronization of the network. Previous simulations have led us to reject values of $b$ corresponding to the Region I of the Fig.~\ref{fig_figura_ISI_b_a_89_prueba}, because in this case, the neuron is highly sensitive to the value of $b$ and even the small mismatch we propose makes almost impossible to have a great synchronization in the network without considering a very high coupling strength between the neurons, which would be unrealistic. We also discard values from the Region II because the results are quite similar to those obtained in the case we propose ($b=0.35$).

With these features in mind, we proceed to simulate the behavior of the previously described small-world network . In our case, we make a network of $50$ neurons ($N=50$), where each neuron is initially connected to its four nearest neighbours ($l=2$). We will increase the rewiring probability $p$ to analyze its effect on the global behavior of the network. Finally,  the model parameters of the neuron are fixed at $a=0.89$, $c=0.28$ and $I=0.03$. These parameters will remain fixed all throughout the paper.

The objective of the numerical simulations is to see how variations in the rest of the model parameters affect the global synchronization of the network and to the firing frequency of the neurons. To analize the synchronization between the neurons of the network, we introduce the order parameter $R$ as in Ref.~\cite{Ojalvo93}. This parameter allows us to measure the synchronization in phase and in amplitude between different signals. In our case, we will apply this parameter to study the synchronization between the variable $x$ of each neuron from the network.  Its mathematical expression is given by:
\begin{equation}\notag
		R(x)=\frac{\left<\overline{x}^2\right>-\left<\overline{x}\right>^2}{\overline{\left<x^2\right>-\left<x\right>^2}},
	\end{equation}\label{Requation}
where $\overline{x}$ is the mean average over all neurons,  and $\left<x\right>$ is the mean in time, i.e., the mean of the variable $x$ of each neuron.  Accordingly, $R(x)\in\left[0,1\right]$, where $R(x)=0$ means total de-synchronization and $R(x)=1$ means total synchronization of the time series $x$.

\begin{figure}[ht!]
\centering
\includegraphics[width=0.9\textwidth]{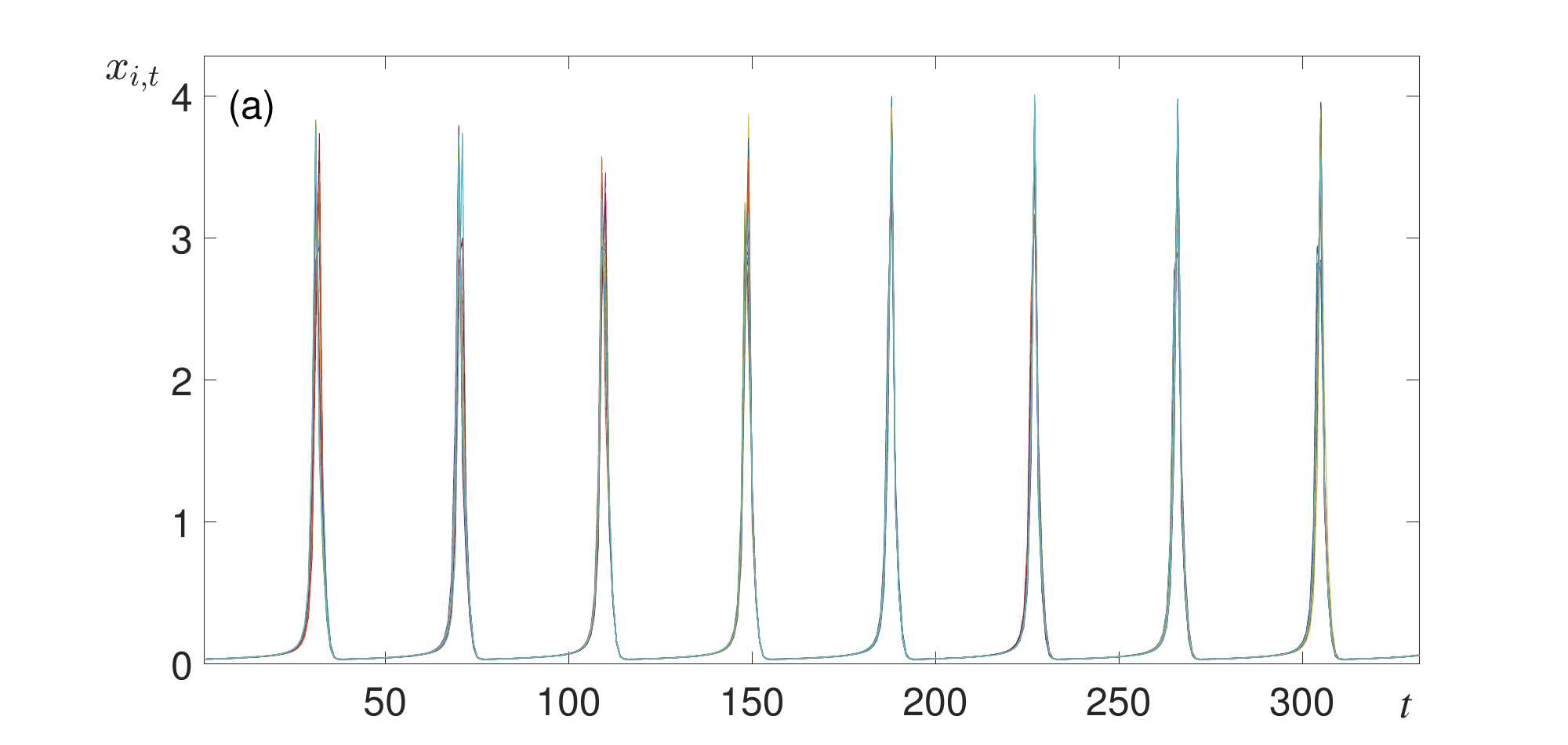}

\includegraphics[width=0.9\textwidth]{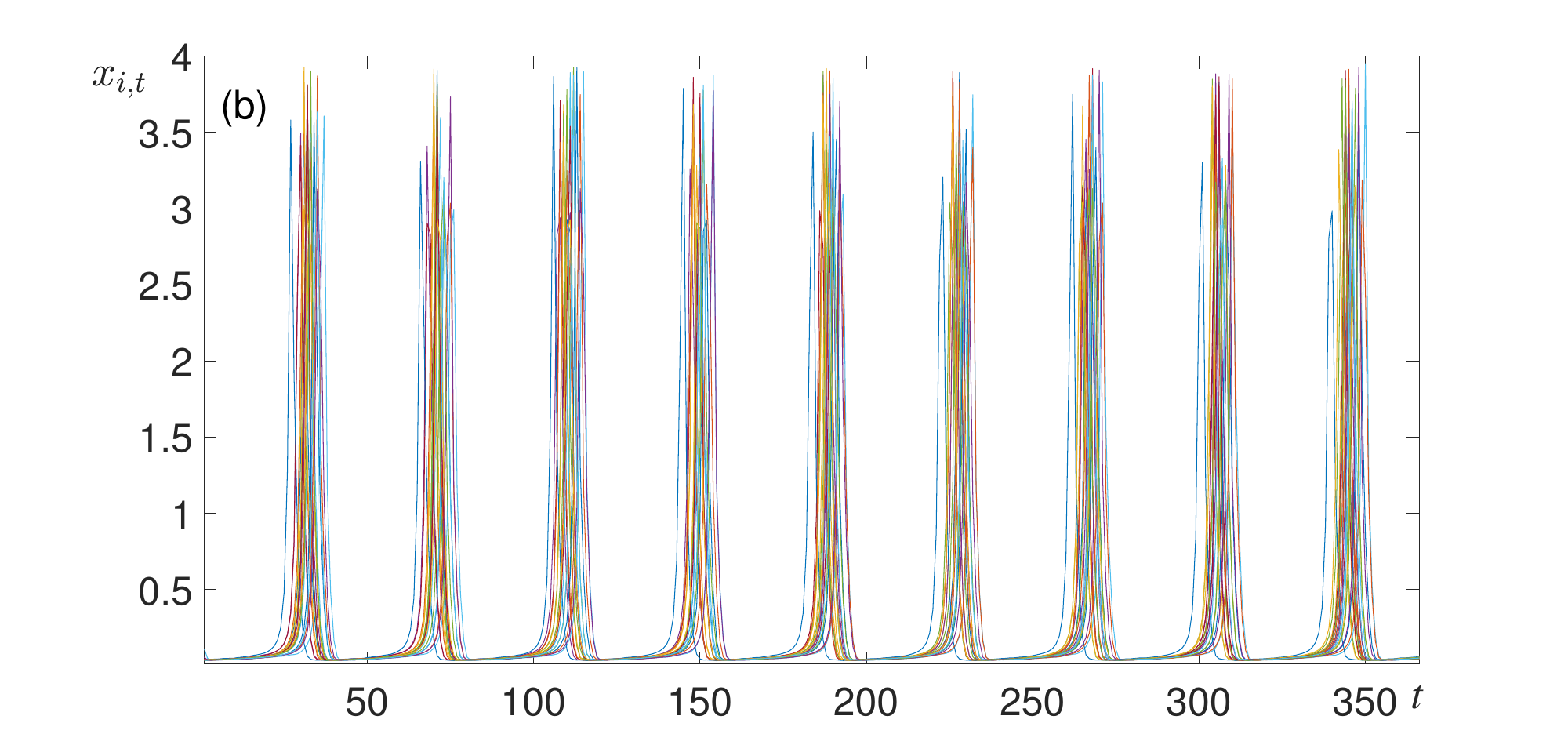}
 \caption{\textbf{The time series of the variable $x_{i,t}$ for the neurons of the network.} In both figures, the model parameters and the noise intensity are the same ($a=0.89$,  $c=0.28$, $I=0.03$ and $\varepsilon=0.003$) and  the values of $b$ for the neurons are randomly chosen from the range $0.3464 \le b \le 0.3535$ ($b\pm 0.01b \xi_{b} $), where $\xi_{b}$ is  uncorrelated white Gaussian noise with parameters $\langle\xi_{b}\rangle=0,\;\langle\xi^2_{b}\rangle=1$, but the coupling strength and the properties of the network are different. The order parameter $R$ is calculated for both cases: in case (a) where the neurons are very well synchronized ($R=0.9716$) and in case (b) where the neurons are unsynchronized ($R=0.210$). (For clarity of the figure only the time series from 10 out of the 50 neurons that are in the network are plotted).}
\label{figura_Rs}
\end{figure}

As an example, we have plotted Fig.~\ref{figura_Rs}, which shows the times series of the variable $x$ from  $10$ neurons under two different working conditions. In this figure, the relationship between the synchronization of the neurons and the value of the order parameter $R$ can be clearly appreciated.

In our case, we will calculate the value of the order parameter $R$ as a function of the noise intensity $\varepsilon$, coupling strength $k$ and the probability of rewiring in the network $p$. Additionally, we will evaluate the parameter $R$ as a function of the ratio between excitatory and inhibitory connections. The number of neurons with a mismatch in the parameter $b$ also plays a crucial role in the synchronization of the network. For this reason, the number of neurons with mismatch in parameter $b$ will be gradually increased.

Every result of this work is obtained as the mean value of $50$ simulations for the same set of parameters values, starting from different initial conditions. Note that in each figure the rewiring probability of the connections in the network, $p$, increases. We start with a network with no rewiring, $p=0$, and finish with a network where the probability of rewiring is $p=0.25$.

\subsubsection{Excitatory coupling in the small-world network}\label{numericalresultsNn_1}

We will start considering an homogeneous neuron network where all the neurons take the same value for the parameter $b=0.35$ and then gradually increase the number of neurons with different values for the parameter $b$.
Besides, the parameter $R$ will be calculated for different values of the rewiring probability, $p$, that will allow us to see how the randomness of the network affects the global synchronization of the system.

\begin{figure}[ht!]
\centering
 \includegraphics[width=\textwidth]{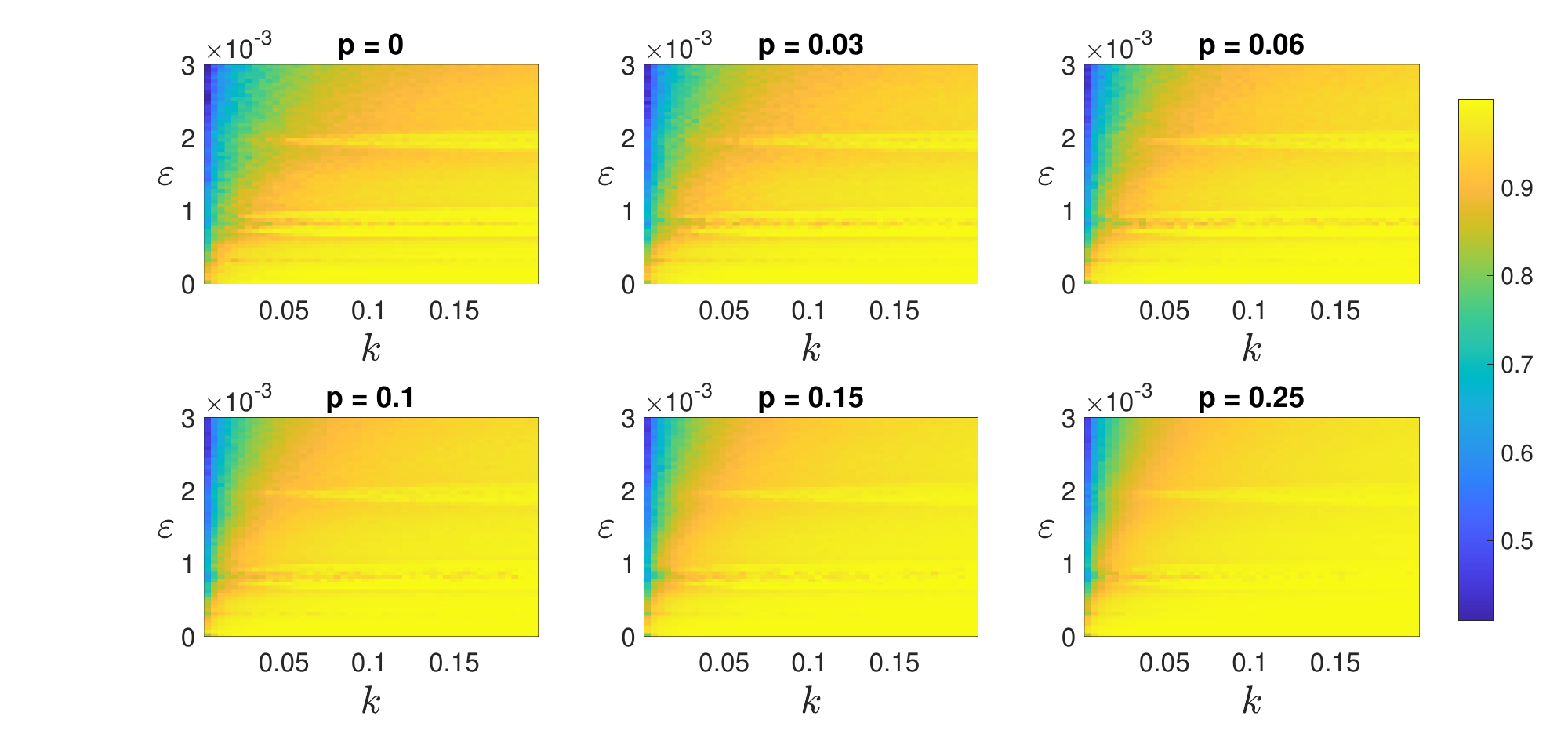}
 \caption{\textbf{Plot of the synchronization order parameter R}. Representation of the parameter $R$ for a network where all neurons have the same values for the parameters, $b=0.35$, $a=0.89$, $c=0.28$ and $I=0.03$.}
 \label{fig_R_b_035_n2_0}
\end{figure}

In Fig.~\ref{fig_R_b_035_n2_0}, where all the neurons are identical, the network is synchronized for almost all the points of the parameter plane $(k,\varepsilon)$ and only for low values of the cloupling strength $k$ and high values of noise intensity $\varepsilon$ does the neuron network become desynchronized. This result is quite expected because the noise intensity makes the signal of the neurons more chaotic, and low values of the coupling strength makes a weak interaction between the neurons of the network, so the influence between them does not affect the synchronization. It can also be appreciated that as the probability of rewiring of the network $p$ increases, the synchronization of the global system also increases. An interesting effect observed in Fig.~\ref{fig_R_b_035_n2_0} is that for all cases of $p$ there are two bands centered around $\varepsilon=0.8\cdot 10^{-3}$ and $\varepsilon=2\cdot 10^{-3}$ where the value or $R$ reaches its highest values regardless of the value of the coupling strength, that is, there is a critical value of the noise intensity that optimizes the global synchronization of the neuron network.

\begin{figure}[ht!]
\centering
 \includegraphics[width=\textwidth]{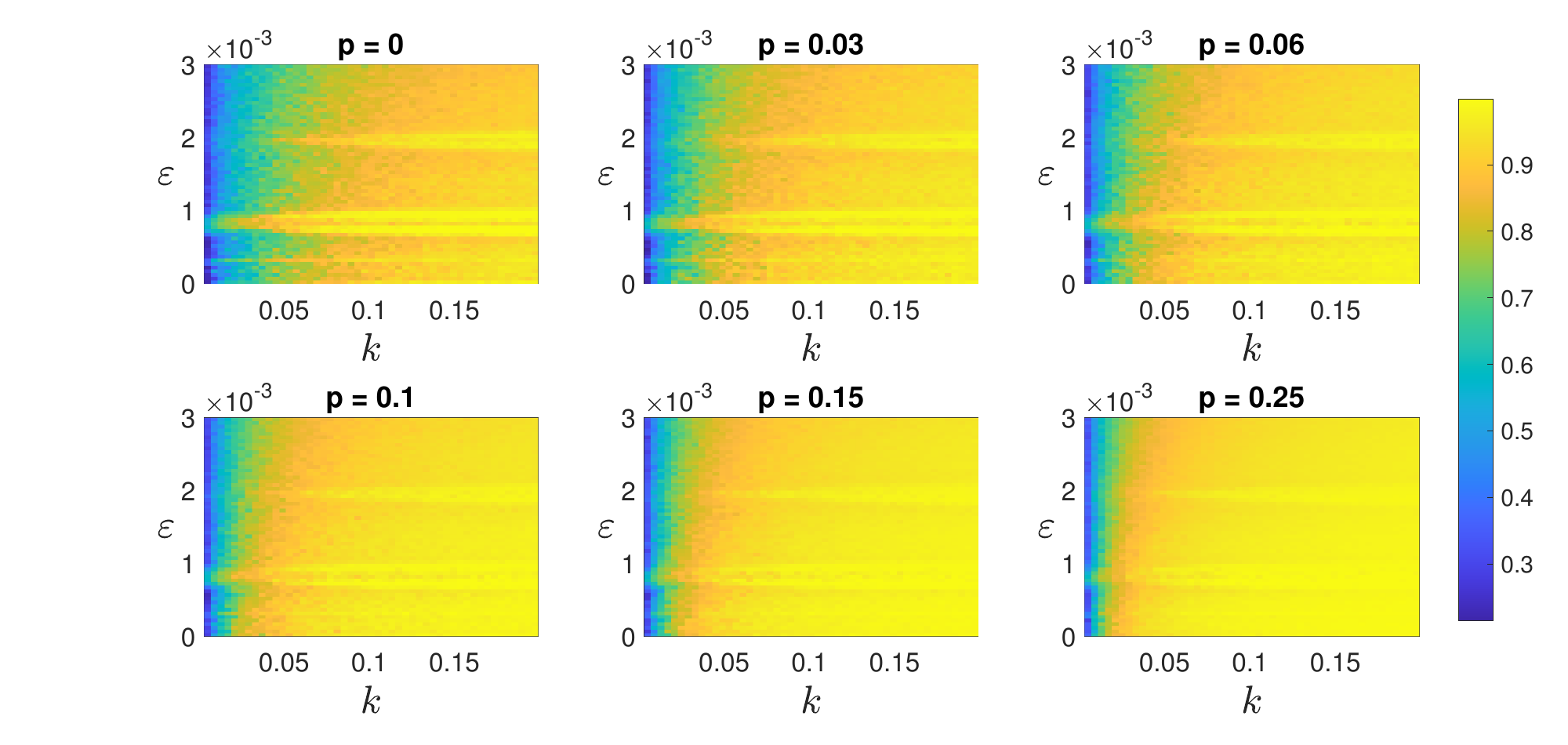}
\caption{\textbf{Plot of the synchronization order parameter R}. Representation of the parameter $R$ for a network where 25 out the 50 neurons have a random value for the parameter $b$. The mean value of the parameter $b=0.35$ and the mismatch is randomly selected with maximum value $\Delta b=0.01b$. The rest of the parameters are fixed for all the neurons, $a=0.89$, $c=0.28$ and $I=0.03$.}
\label{fig_R_b_035_n2_25}
\end{figure}

\begin{figure}[ht!]
\centering
 \includegraphics[width=\textwidth]{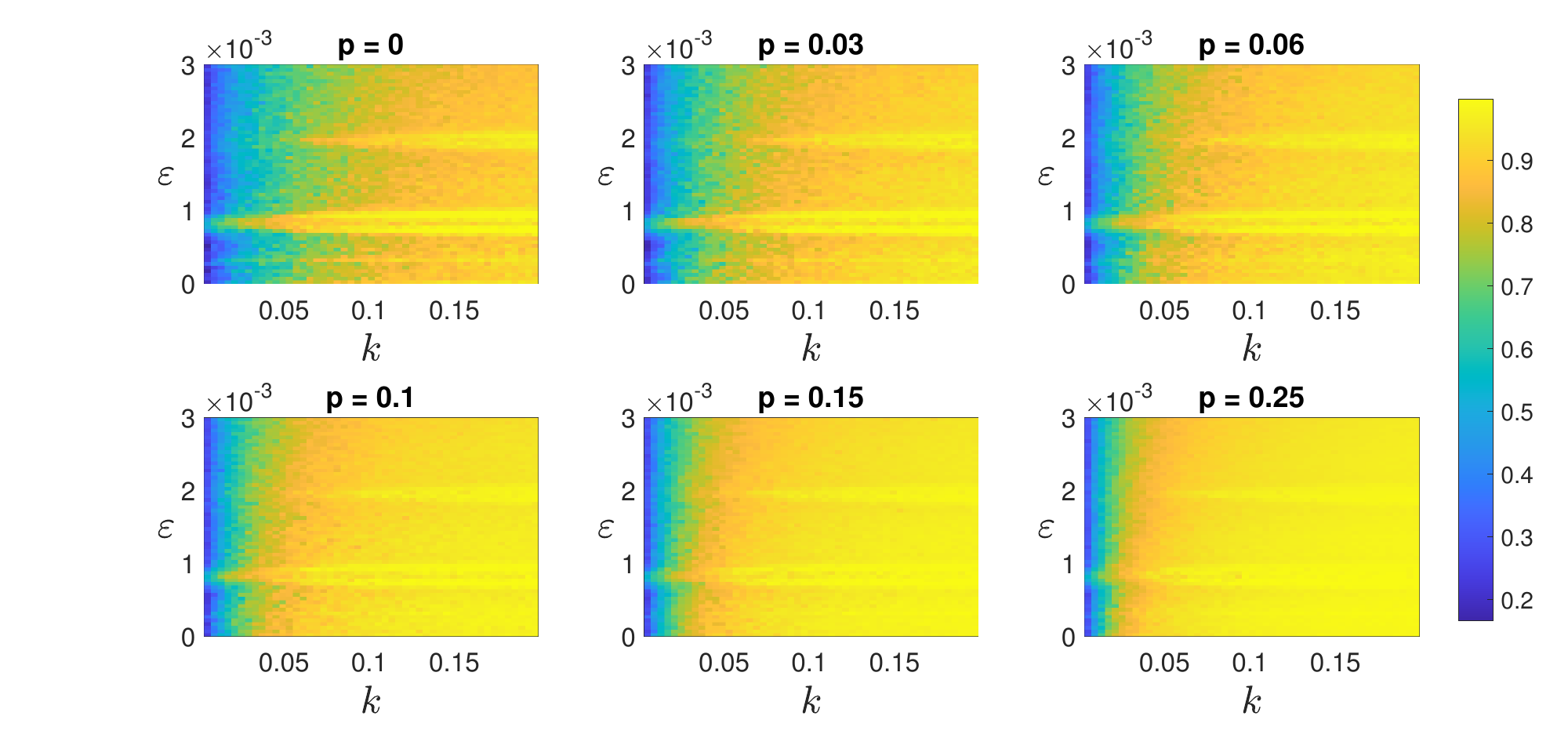}
\caption{\textbf{Plot of the synchronization order parameter R}. Representation of the order parameter $R$ for a network where all neurons of the network have a random value for the parameter $b$. The mean value of the parameter $b=0.35$ and the mismatch is randomly selected with maximum value $\Delta b=0.01b$. The rest of the parameters are fixed for all the neurons, $a=0.89$, $c=0.28$ and $I=0.03$.
}
\label{fig_R_b_035_n2_50}
\end{figure}

In Fig.~\ref{fig_R_b_035_n2_25}, $25$ out of the $50$ neurons have a mismatch in the parameter $b$ and in Fig.~\ref{fig_R_b_035_n2_50}, all the neurons have this mismatch, that is, all neurons are different. In these cases we can observe similar phenomena to those observed in the homogenous case: increasing the rewiring probability of the network $p$ enhances the synchronization, and the bands with the highest synchronization are also reproduced, centered around the same values of the $\varepsilon$. Finally, it can be appreciated that increasing the heterogeneity of the neurons worsens the network synchronization: the region with low values of the parameter $R$ increases from Fig.~\ref{fig_R_b_035_n2_0} to Fig.~\ref{fig_R_b_035_n2_50}.

\begin{figure}[ht!]
\centering
 \includegraphics[width=\textwidth]{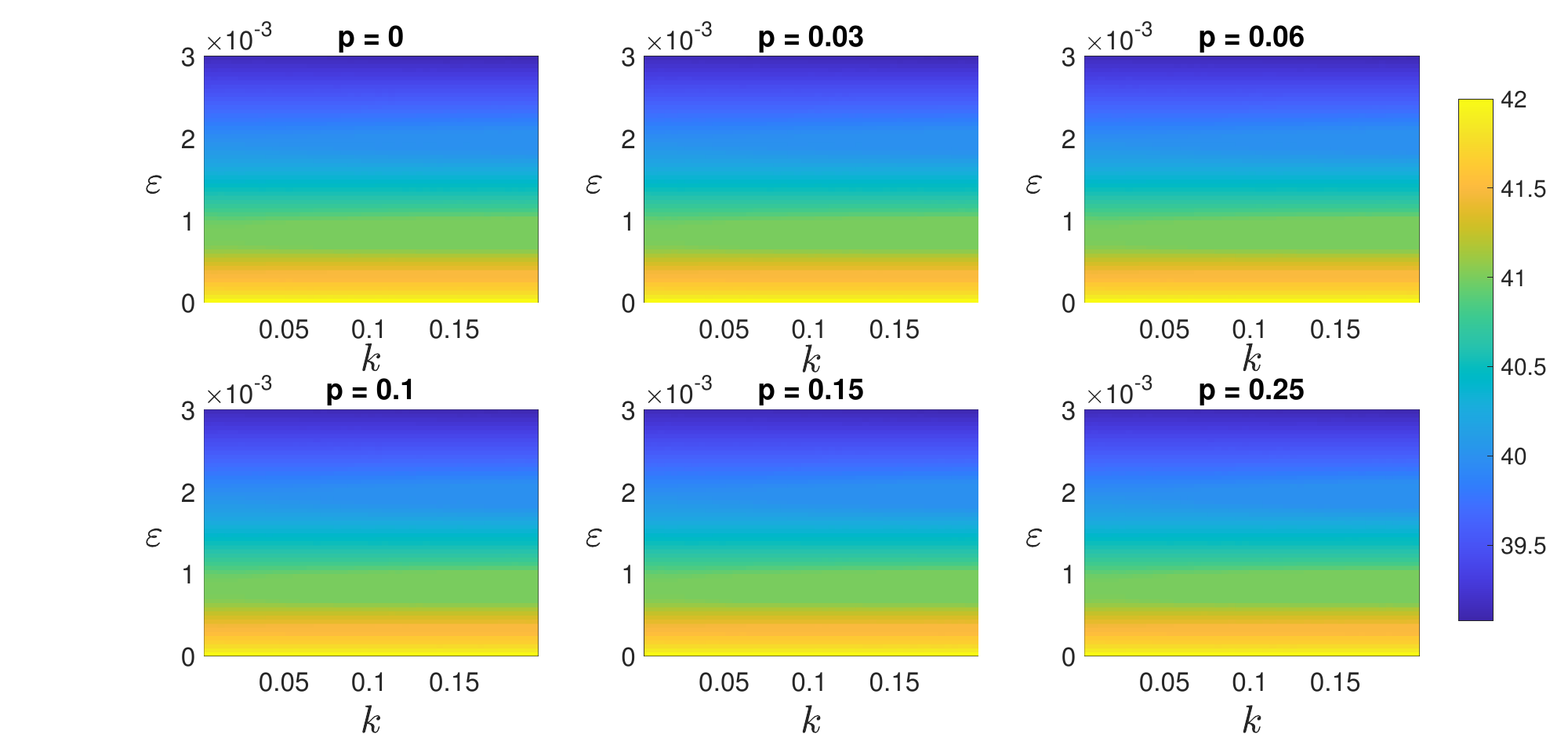}
 \caption{\textbf{Plot of the inter-spike interval ($ISI$)}. Representation of the mean value of the $ISI$ for a network where all neurons have the same values for the parameters, $b=0.35$, $a=0.89$, $c=0.28$ and $I=0.03$. }\label{fig_ISI_b_035_n2_0}
\end{figure}

\begin{figure}[ht!]
\centering
 \includegraphics[width=\textwidth]{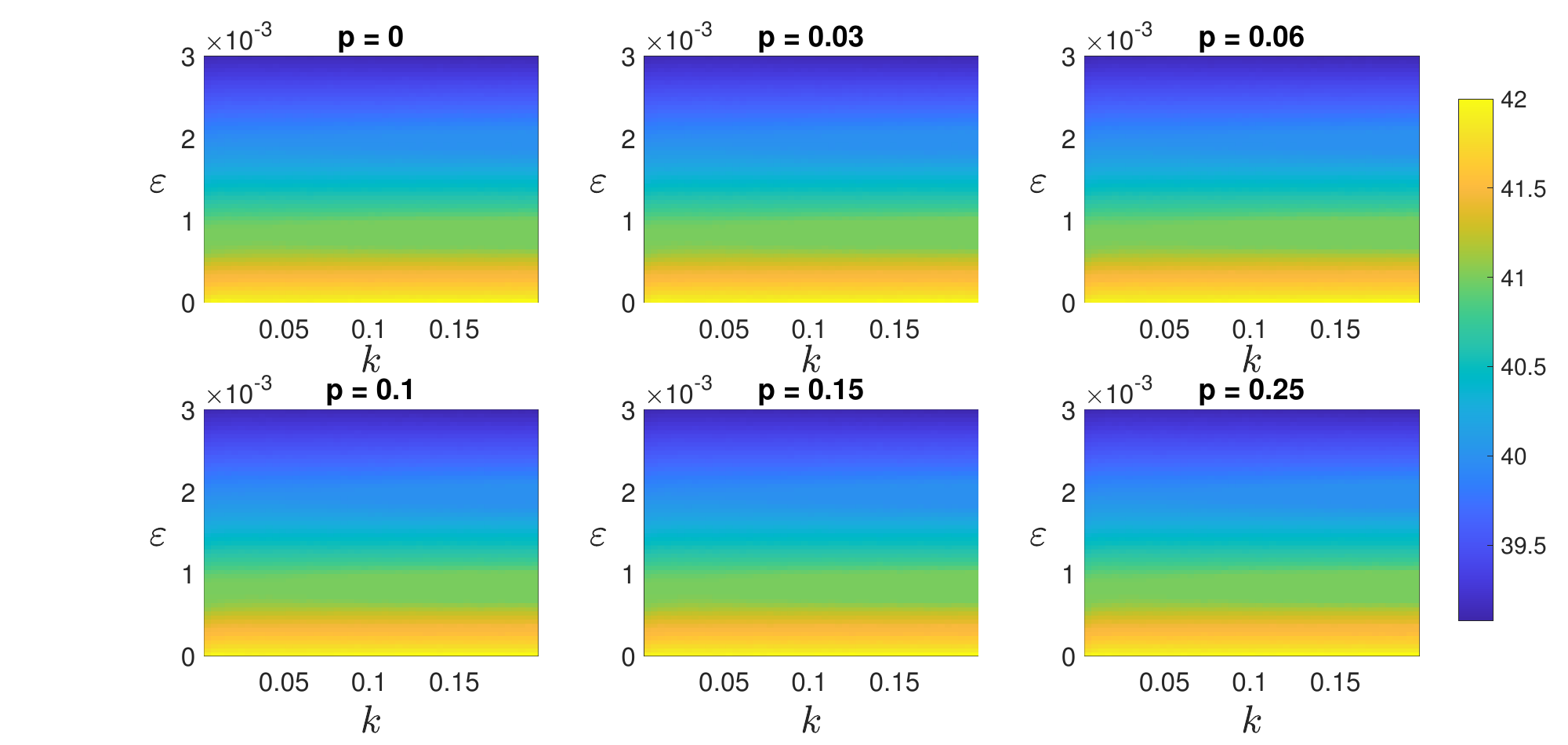}
\caption{\textbf{Plot of the inter-spike interval ($ISI$)}. Representation of the mean value of the $ISI$ for a network where 25 out of the 50 neurons have a random value for the parameter $b$. The mean value of the parameter $b=0.35$ and the mismatch is randomly selected with maximum value $\Delta b=0.01b$. The rest of the parameters are equal for all the neurons, $a=0.89$, $c=0.28$ and $I=0.03$.}\label{fig_ISI_b_035_n2_25}
\end{figure}

\begin{figure}[ht!]
\centering
 \includegraphics[width=\textwidth]{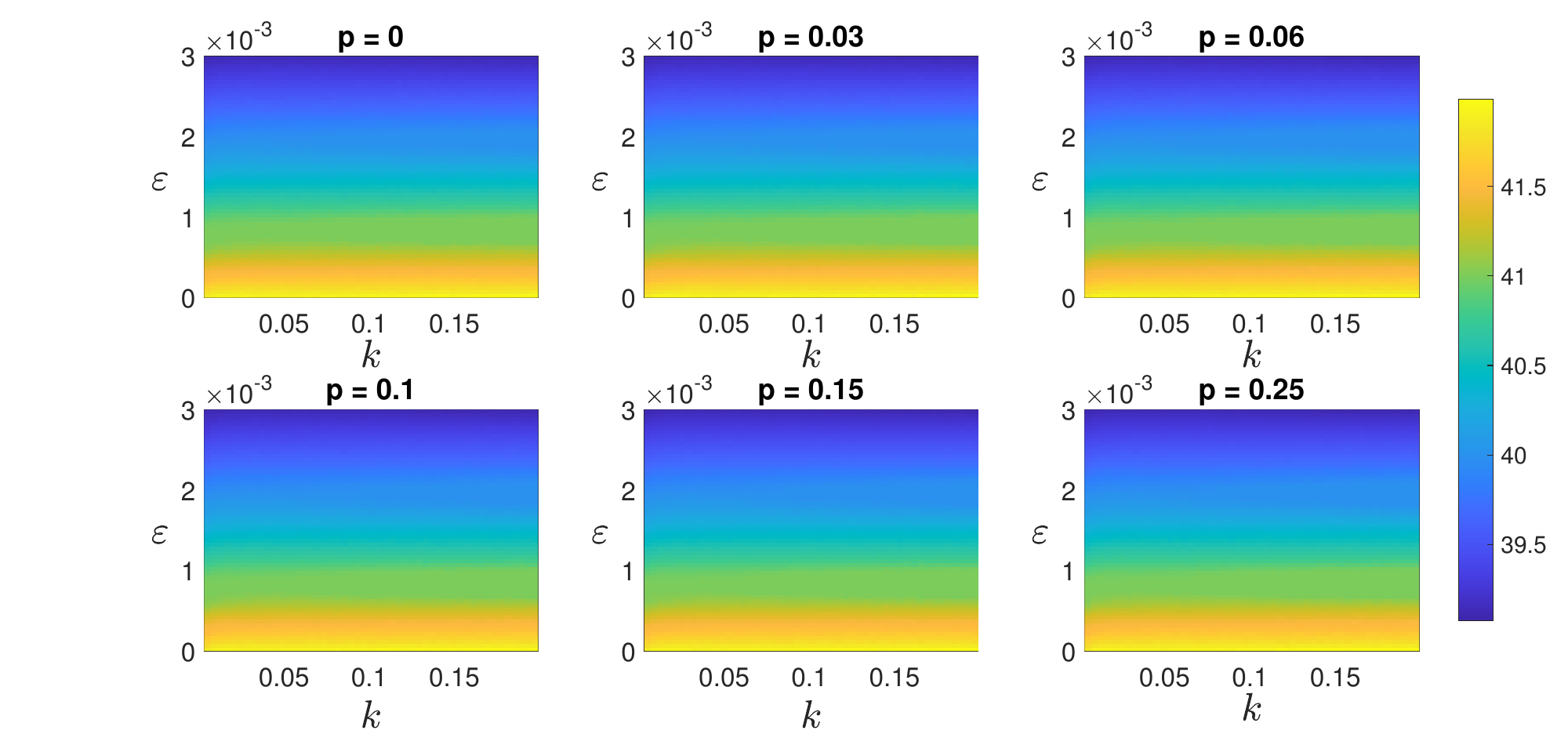}
\caption{\textbf{Plot of the inter-spike interval ($ISI$)}. Representation of the mean value of the $ISI$ for a network where all the 50 neurons of the network have a random value for the parameter $b$. The mean value of the parameter $b=0.35$ and the mismatch is randomly selected with maximum value $\Delta b=0.01b$.}\label{fig_ISI_b_035_n2_50}
\end{figure}

As mentioned in the introduction, the firing frequency is of special interest because it is related to neural coding. For this reason, the value of the inter-spike interval has been studied as a function of the noise intensity and the coupling strength. When we refer to parameter $ISI$, we are representing the mean value of the inter-spike interval of every neuron in the network. The $ISI$ is also calculated for different values of the rewiring probability of the network $p$. As in the previous study of the order parameter $R$, the value of the parameter $ISI$ has been obtained for three different cases depending on the number of neurons with a mismatch in the parameter $b$. In Fig.~\ref{fig_ISI_b_035_n2_0} all the neurons are identical, in Fig.~\ref{fig_ISI_b_035_n2_25}, $25$ out the $50$ neurons have mismatch in the parameter $b$ and finally in Fig.~\ref{fig_ISI_b_035_n2_50}, all the neurons have a different value for the parameter $b$.

If we compare the figures obtained, we can observe that all the figures are quite similar, in fact, the differences between Fig.~\ref{fig_ISI_b_035_n2_0}  and Fig.~\ref{fig_ISI_b_035_n2_25} are almost negligible, and for Fig.~\ref{fig_ISI_b_035_n2_50} what we observe is that the range of values of the $ISI$ is slightly reduced compared to the one obtained for Figs.~\ref{fig_ISI_b_035_n2_0}-\ref{fig_ISI_b_035_n2_25}. In all cases, it can be observed how the $ISI$ decreases as the noise intensity $\varepsilon$ increases and that there is no direct relationship between the $ISI$ and the coupling strength $k$ or the rewiring probability in the network $p$. In all cases, we have also calculated the standard deviation in the $ISI$ and we have obtained that its value is very close to zero, even in the regions of the parameter plane $(k,\varepsilon)$ where the parameter $R$ takes its maximum values, we can assure that is zero. That means that all the neurons have the same firing frequency, and the neuron network is very stable in the frequency domain.

\subsubsection{Inhibitory coupling in the small-world network}\label{numericalresultsNn_2}

Normally, when two neurons are connected by inhibitory coupling, their synchronization worsens. But can this observation be extended to a neural network? Can we assume that the introduction of some inhibitory connections must desynchronize the system? To answer these questions, we propose to introduce a percentage of inhibitory connections in the network and analyze its effect on global synchronization. The percentage of inhibitory coupling has a great dependence on the area of the brain analyzed and on the type of neurons connected \cite{attila99}, and its effect is quite important in the global behavior of the neuron network  \cite{Borgers03, Borgers05}. As in the previous section, the figures obtained for the cases corresponding to values of $b$ corresponding to the Regions II and III remarked in Fig.~\ref{fig_figura_subplot_a_089_b_noise} are quite similar, so we focus our simulations on the case $b=0.35$. The procedure is similar to the excitatory case: we will start by considering that all the neurons have the same value for the parameter $b$ and gradually increase the number of neurons with a mismatch in this parameter.

We will start considering that a $1\%$ of the total links in the network are inhibitory connections.

\begin{figure}[ht!]
\centering
 \includegraphics[width=\textwidth]{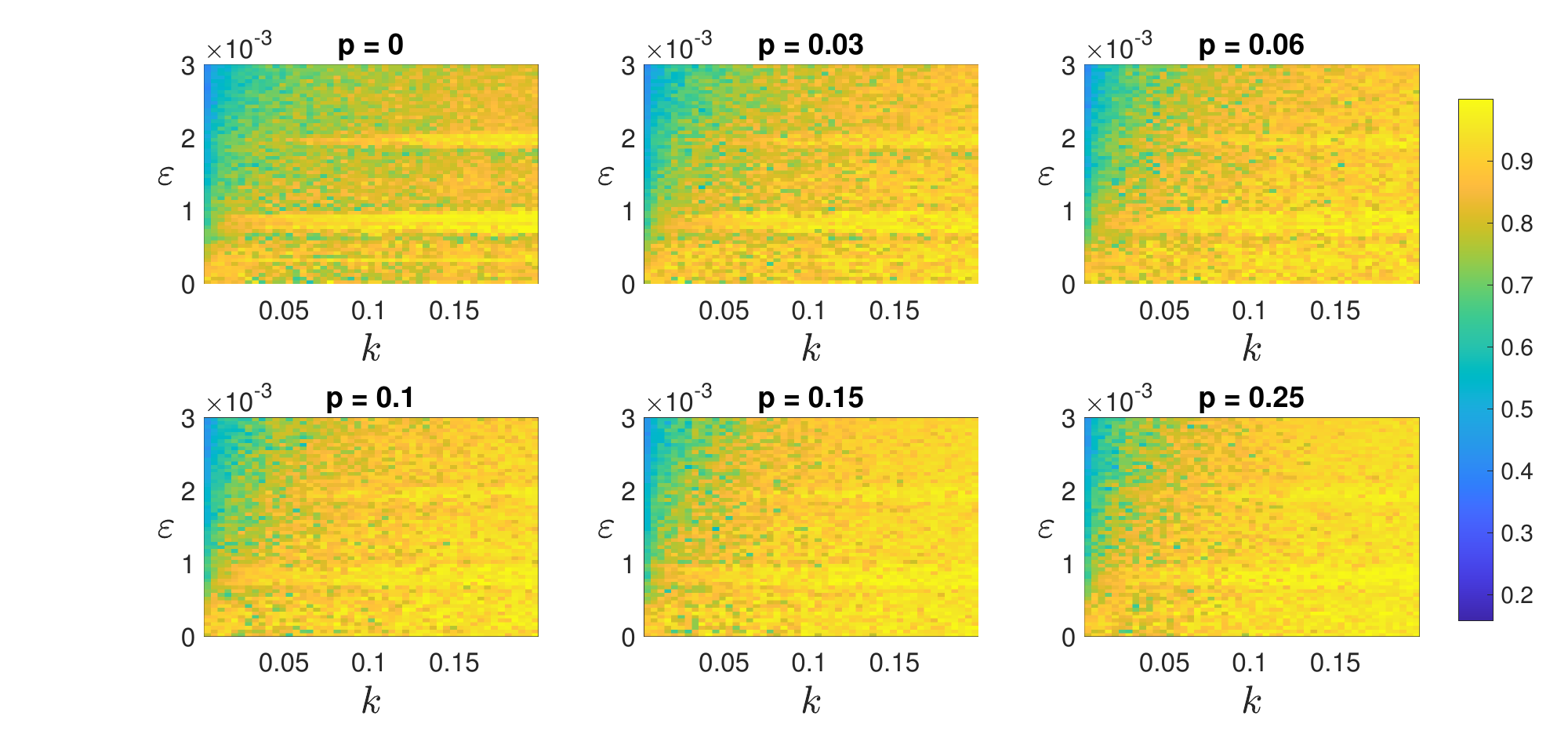}
 \caption{\textbf{Plot of the synchronization order parameter R}. Representation of the order parameter $R$ for a network where all neurons have the same values for the parameters, $b=0.35$, $a=0.89$, $c=0.28$ and $I=0.03$. In the network an $1\%$ of the connections are inhibitory couplings.}\label{fig_R_b_035_n2_0_inhi_01}
\end{figure}

\begin{figure}[ht!]
\centering
 \includegraphics[width=\textwidth]{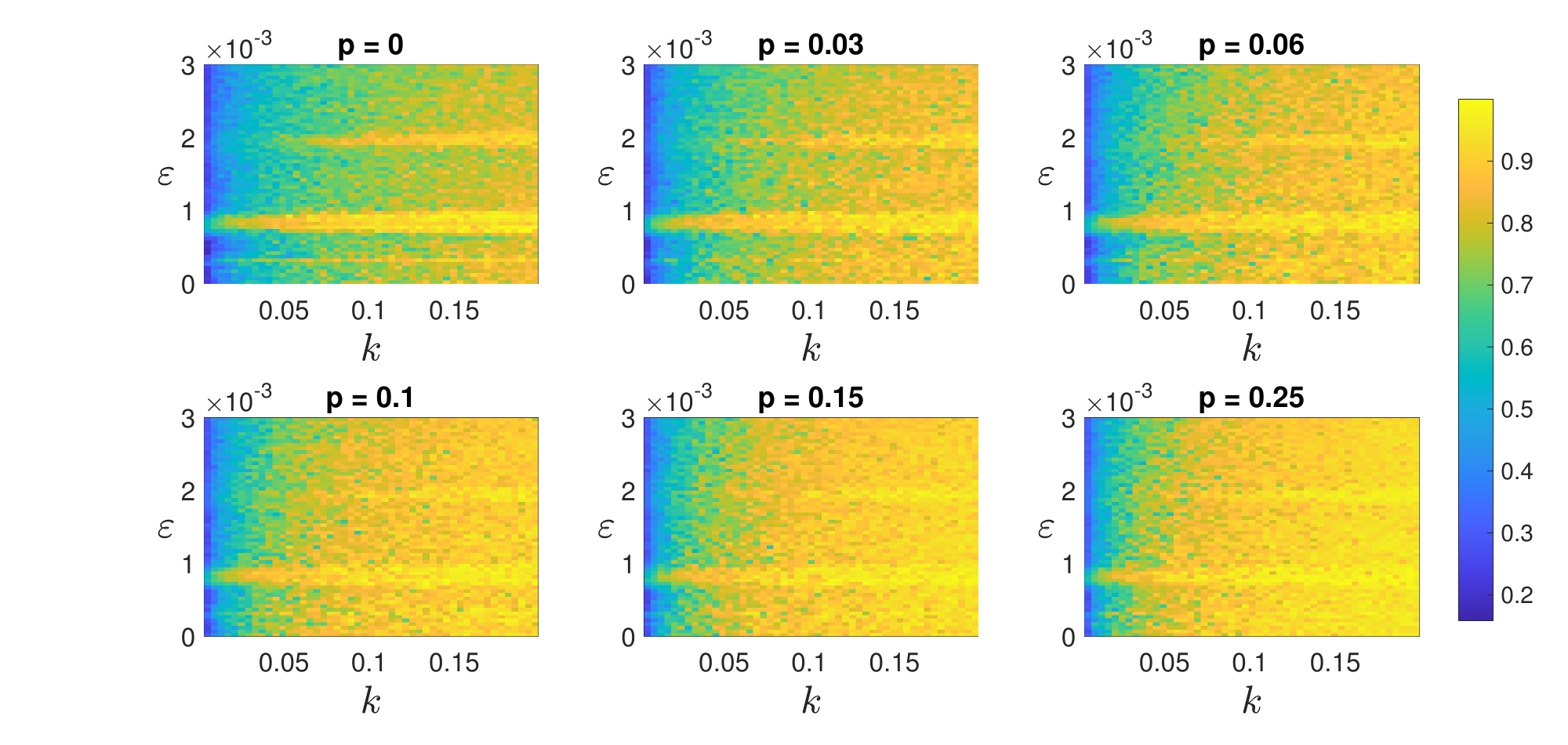}
\caption{\textbf{Plot of the synchronization order parameter R}. Representation of the order parameter $R$ for a network where 25 out of the 50 neurons have a random value for the parameter $b$. The mean value of the parameter $b=0.35$ and the mismatch is randomly selected with maximum value $\Delta b=0.01b$. The rest of the parameters are equal for all the neurons, $a=0.89$, $c=0.28$ and $I=0.03$. In the network an $1\%$ of the connections are inhibitory couplings.}
\label{fig_R_b_035_n2_25_inhi_01}
\end{figure}

\begin{figure}[ht!]
\centering
 \includegraphics[width=\textwidth]{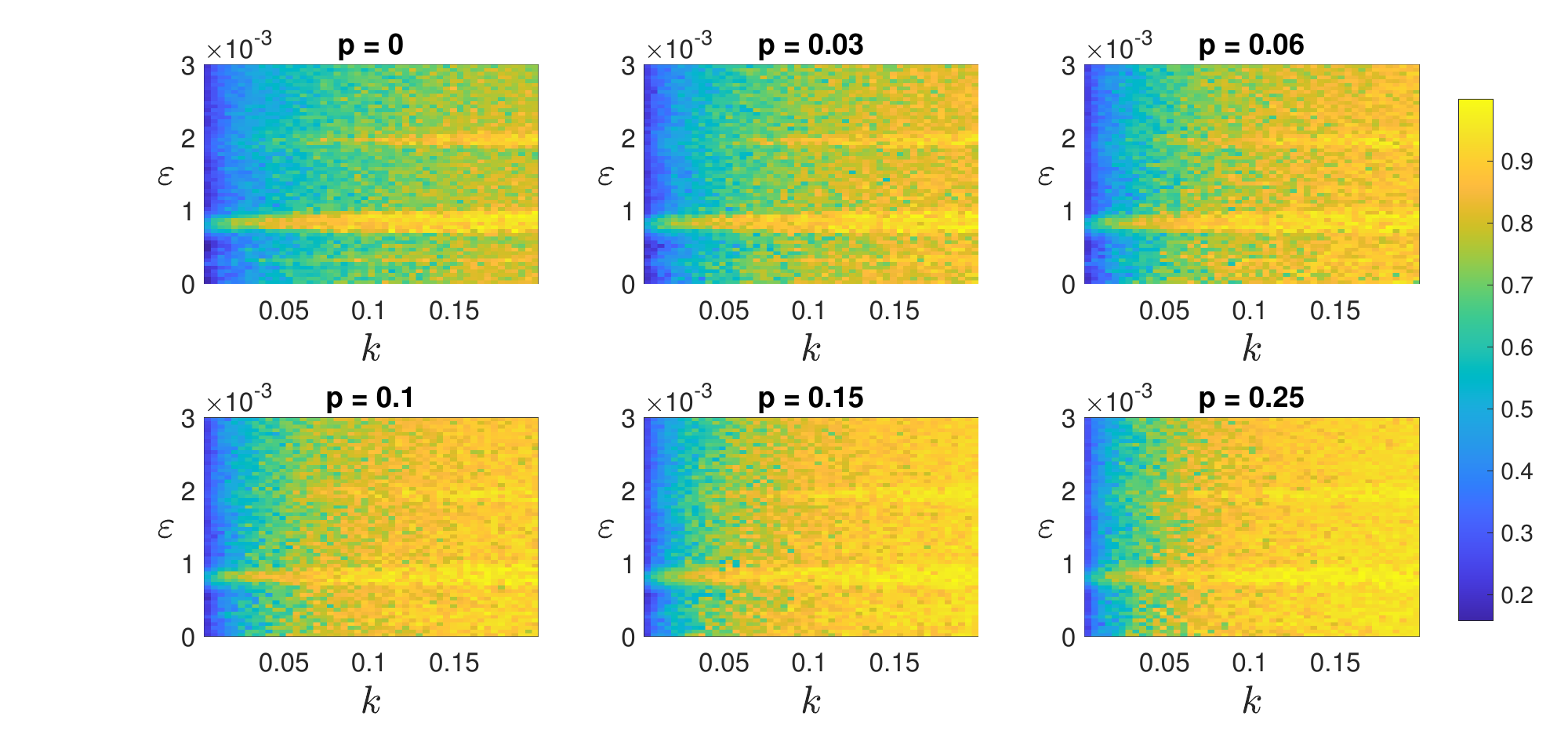}
\caption{\textbf{Plot of the synchronization order parameter R}. Representation of the order parameter $R$ for a network where all the 50 neurons from the network have a random value for the parameter $b$. The mean value of the parameter $b=0.35$ and the mismatch is randomly selected with maximum value $\Delta b=0.01b$. The rest of the parameters are equal for all the neurons, $a=0.89$, $c=0.28$ and $I=0.03$. In the network an $1\%$ of the connections are inhibitory couplings.}
\label{fig_R_b_035_n2_50_inhi_01}
\end{figure}

The two main results that we can extract from these figures are: (i) In all cases, from the homogeneous case (Fig.~\ref{fig_R_b_035_n2_0_inhi_01}) to the case where all the neurons have a mismatch in the parameter $b$, (Fig.~\ref{fig_R_b_035_n2_50_inhi_01}), the increase of the rewiring probability $p$ produces better synchronization of the network. It can be clearly appreciated how the area with high values for the parameter $R$ increases as $p$ increases. Additionally, the two bands centered around $\varepsilon=0.8\cdot 10^{-3}$ and $\varepsilon=2\cdot 10^{-3}$, observed for the case where all connections are excitatory and where the network reached its maximum values of synchronization, are also observable for these cases.   (ii) The increase of the number of neurons with mismatch in the parameter $b$ results in worse synchronization of the network. This can be observed by comparing the evolution from Fig.~\ref{fig_R_b_035_n2_0_inhi_01} to Fig.~\ref{fig_R_b_035_n2_50_inhi_01}.

As in the previous case where all the connections were excitatory, we also represent the mean value of the $ISI$ for three different cases depending on the number of neurons with mismatch in the parameter $b$.

\begin{figure}[ht!]
\centering
 \includegraphics[width=\textwidth]{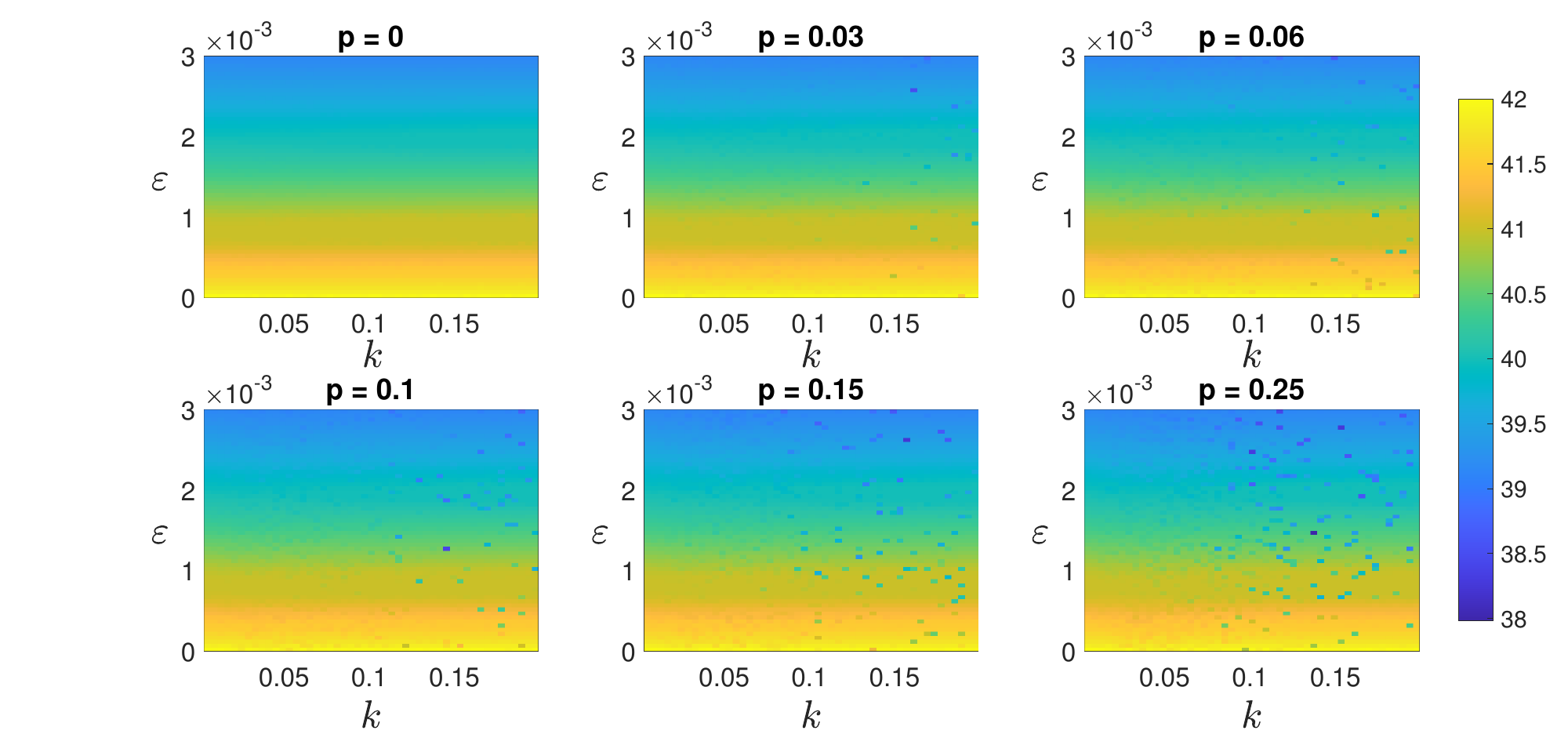}
\caption{\textbf{Plot of the inter-spike interval ($ISI$)}. Representation of the mean value of the $ISI$ for a network where all neurons have the same values for the parameters, $b=0.35$, $a=0.89$, $c=0.28$ and $I=0.03$. In the network an $1\%$ of the connections are inhibitory couplings.}\label{fig_ISI_b_035_n2_0_inhi_01}
\end{figure}

\begin{figure}[ht!]
\centering
 \includegraphics[width=\textwidth]{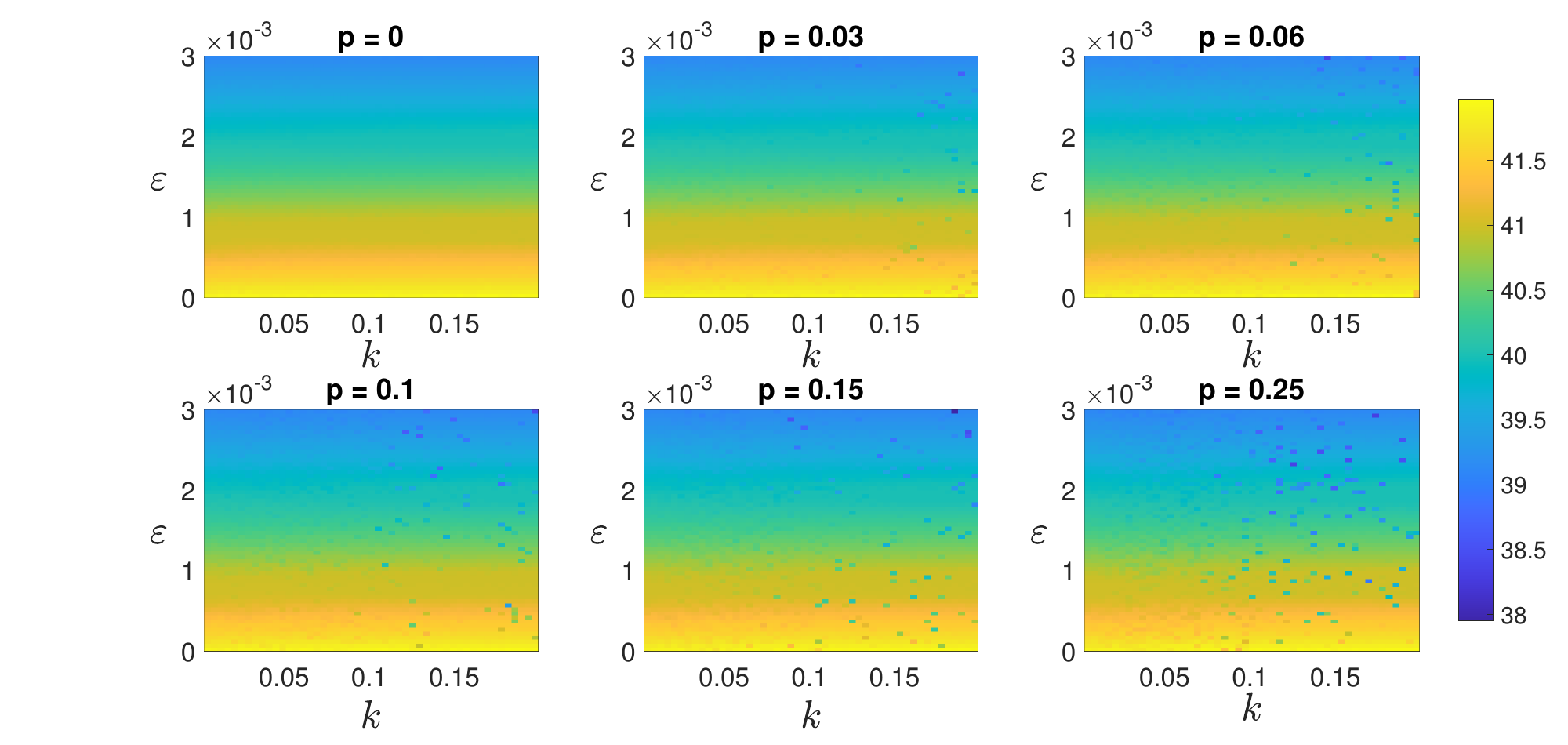}
\caption{\textbf{Plot of the inter-spike interval ($ISI$)}. Representation of the mean value of the $ISI$ for a network where 25 out of the 50 neurons have a random value for the parameter $b$. The mean value of the parameter $b=0.35$ and the mismatch is randomly selected with maximum value $\Delta b=0.01b$. The rest of the parameters are equal for all the neurons, $a=0.89$, $c=0.28$ and $I=0.03$. In the network an $1\%$ of the connections are inhibitory couplings.}
\label{fig_ISI_b_035_n2_25_inhi_01}
\end{figure}

\begin{figure}[ht!]
\centering
 \includegraphics[width=\textwidth]{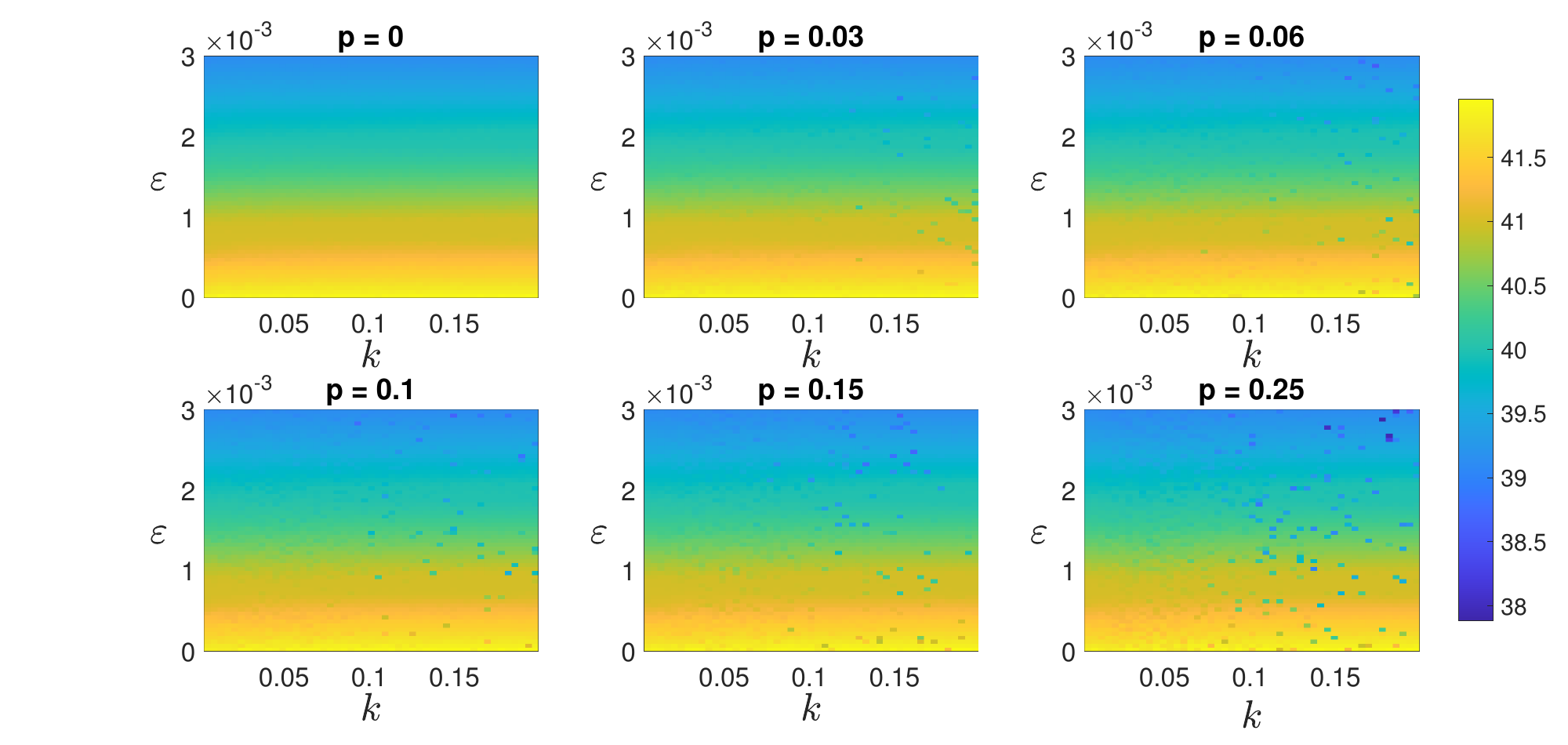}
\caption{\textbf{Plot of the inter-spike interval ($ISI$)}. Representation of the mean value of the $ISI$ for a network where all the 50 neurons from the network have a random value for the parameter $b$. The mean value of the parameter $b=0.35$ and the mismatch is randomly selected with maximum value $\Delta b=0.01b$. The rest of the parameters are equal for all the neurons, $a=0.89$, $c=0.28$ and $I=0.03$. In the network an $1\%$ of the connections are inhibitory couplings.}
\label{fig_ISI_b_035_n2_50_inhi_01}
\end{figure}

In general, upon comparing Figs.~\ref{fig_ISI_b_035_n2_0_inhi_01}-\ref{fig_ISI_b_035_n2_50_inhi_01} we observe that the results are quite similar. In each of these figures, the value of the $ISI$ decreases as the noise intensity $\varepsilon$ increases, but remains constant for all the values of the coupling strength $k$. It can also be appreciated that the range of values of the $ISI$ is slightly lower as the number of neurons with mismatch in the parameter $b$ increases. Finally, we observe that as the number of neurons with a mismatch increases and the rewiring probability in the network $p$ increases, the value of the $ISI$ becomes less stable. That is, in the  subplots of the Figs.~\ref{fig_ISI_b_035_n2_0_inhi_01} -\ref{fig_ISI_b_035_n2_50_inhi_01} corresponding to $p=0.1$, $p=0.15$ and $p=0.25$, the value of the $ISI$ for some points changes significantly compared to the values obtained in the surrounding points. Since the value of the $ISI$ is obtained as the mean value of 50 simulations where the inhibitory links are chosen randomly from all the connections in the network, this leads us to think that there could be a relationship between the $ISI$ and the internal structure of the network.

Finally, we increase the percentage inhibitory connections in the network to a $5\%$. Under this condition, we study the synchronization of the network and the value of the $ISI$ with a similar procedure to the one we did previously.

\begin{figure}[ht!]
\centering
 \includegraphics[width=\textwidth]{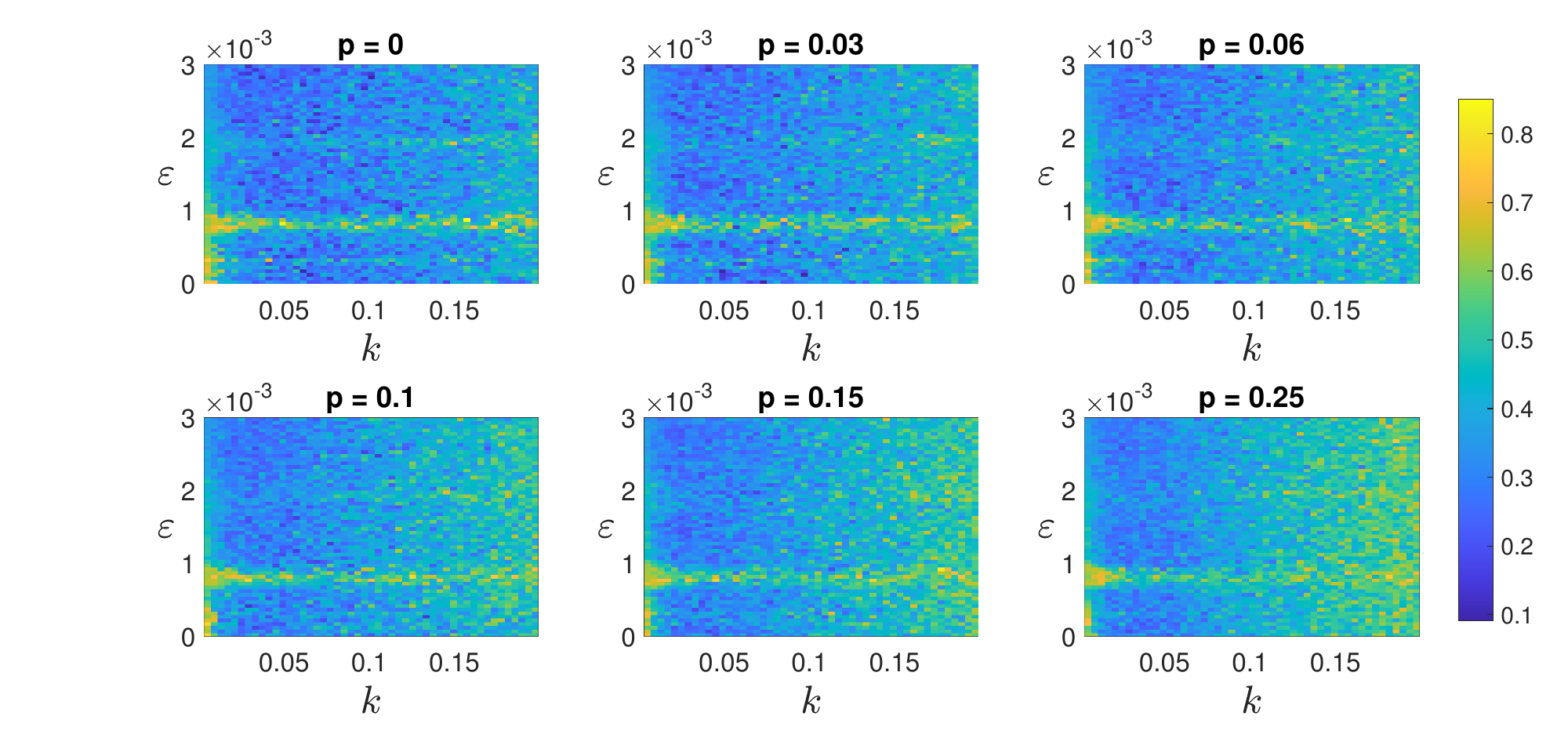}
 \caption{\textbf{Plot of the synchronization order parameter R}. Representation of the order parameter $R$ for a network where all neurons have the same values for the parameters, $b=0.35$, $a=0.89$, $c=0.28$ and $I=0.03$. In the network an $5\%$ of the connections are inhibitory couplings.}
 \label{fig_R_b_035_n2_0_inhi_05}
\end{figure}

\begin{figure}[ht!]
\centering
 \includegraphics[width=\textwidth]{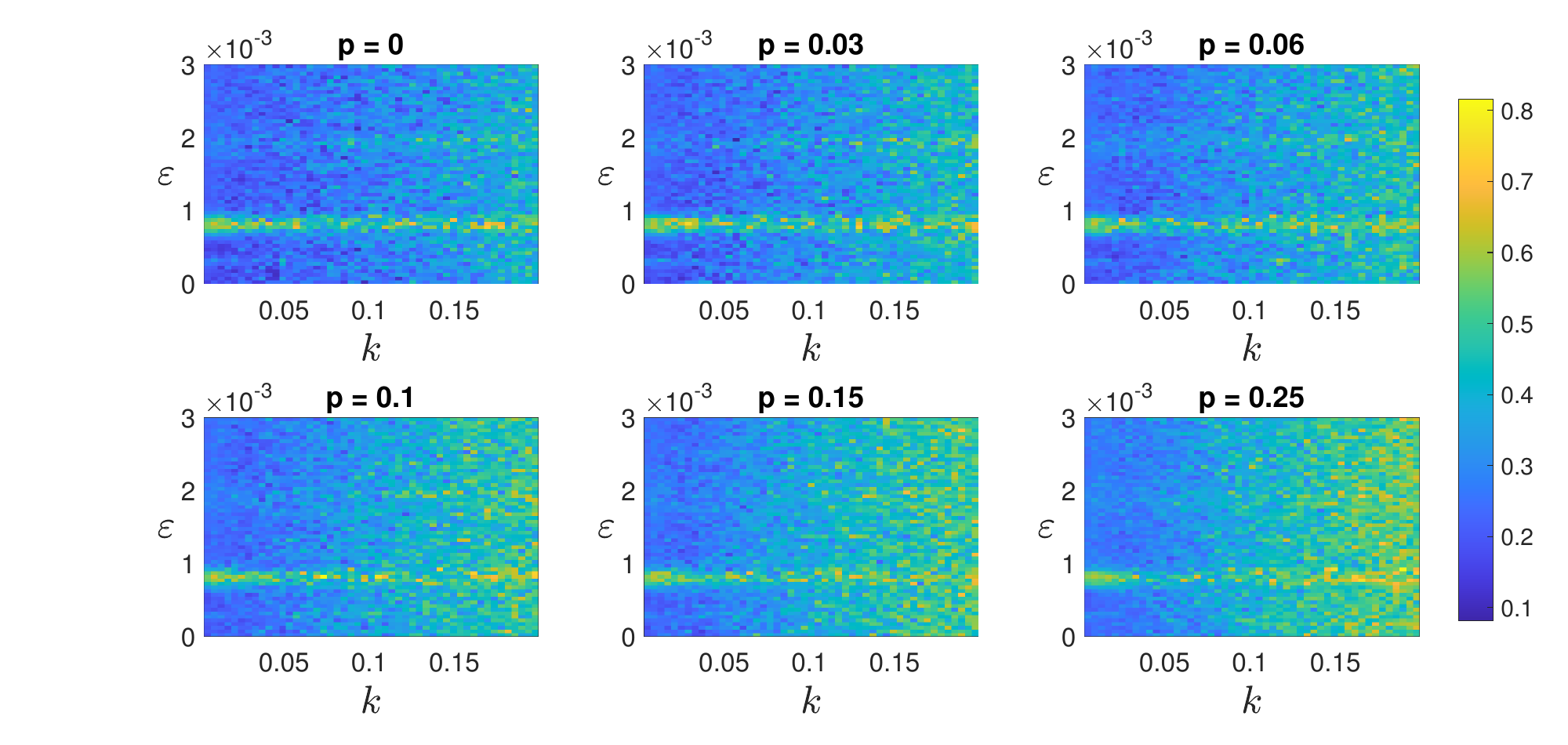}
\caption{\textbf{Plot of the synchronization order parameter R}. Representation of the order parameter $R$ for a network where 25 out of the 50 neurons have a random value for the parameter $b$. The mean value of the parameter $b=0.35$ and the mismatch is randomly selected with maximum value $\Delta b=0.01b$. The rest of the parameters are equal for all the neurons, $a=0.89$, $c=0.28$ and $I=0.03$. In the network an $5\%$ of the connections are inhibitory couplings.}
\label{fig_R_b_035_n2_25_inhi_05}
\end{figure}

\begin{figure}[ht!]
\centering
 \includegraphics[width=\textwidth]{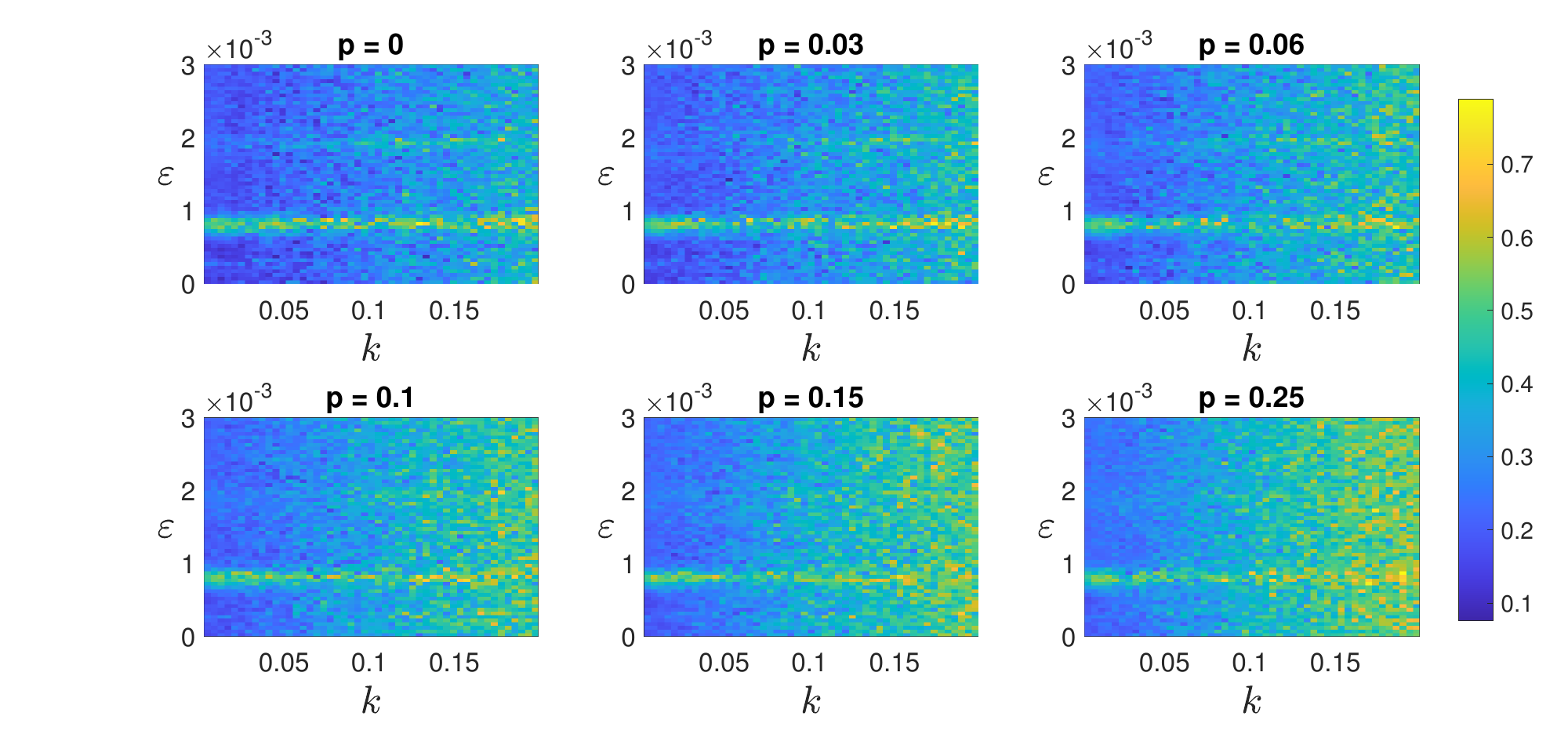}
\caption{\textbf{Plot of the synchronization order parameter R}. Representation of the order parameter $R$ for a network where all the 50 neurons from the network have a random value for the parameter $b$. The mean value of the parameter $b=0.35$ and the mismatch is randomly selected with maximum value $\Delta b=0.01b$. The rest of the parameters are equal for all the neurons, $a=0.89$, $c=0.28$ and $I=0.03$. In the network a $5\%$ of the connections are inhibitory couplings.}
\label{fig_R_b_035_n2_50_inhi_05}
\end{figure}

The results of the study of the synchronization of the network are shown in Figs.~\ref{fig_R_b_035_n2_0_inhi_05}, \ref{fig_R_b_035_n2_25_inhi_05} and \ref{fig_R_b_035_n2_50_inhi_05}. The  first observation is that the value of the order parameter $R$ is significantly lower than that obtained in the previous cases. It is evident how the increase in the percentage of inhibitory connections negatively affects the global synchronization of the network. In all cases, it is noticeable how the increase in the rewiring probability enhances the global synchronization of the network. This effect warrants detailed analysis and could be a promising avenue for further investigation.

\begin{figure}[ht!]
\centering
\includegraphics[width=\textwidth]{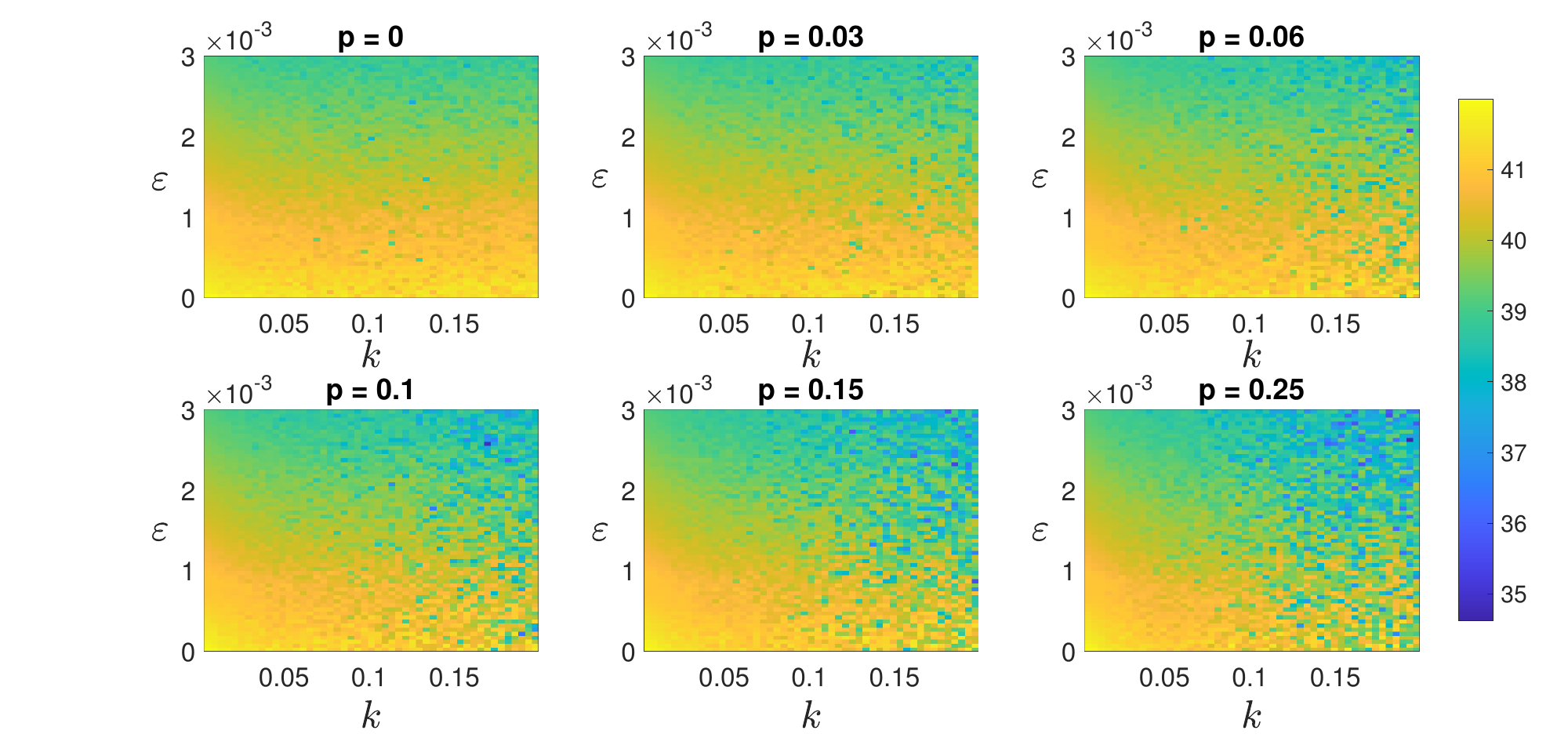}
\caption{\textbf{Plot of the inter-spike interval ($ISI$)}. Representation of the mean value of the $ISI$ for a network where all neurons have the same values for the parameters, $b=0.35$, $a=0.89$, $c=0.28$ and $I=0.03$. In the network a $5\%$ of the connections are inhibitory couplings.}\label{fig_ISI_b_035_n2_0_inhi_05}
\end{figure}

\begin{figure}[ht!]
\centering
 \includegraphics[width=\textwidth]{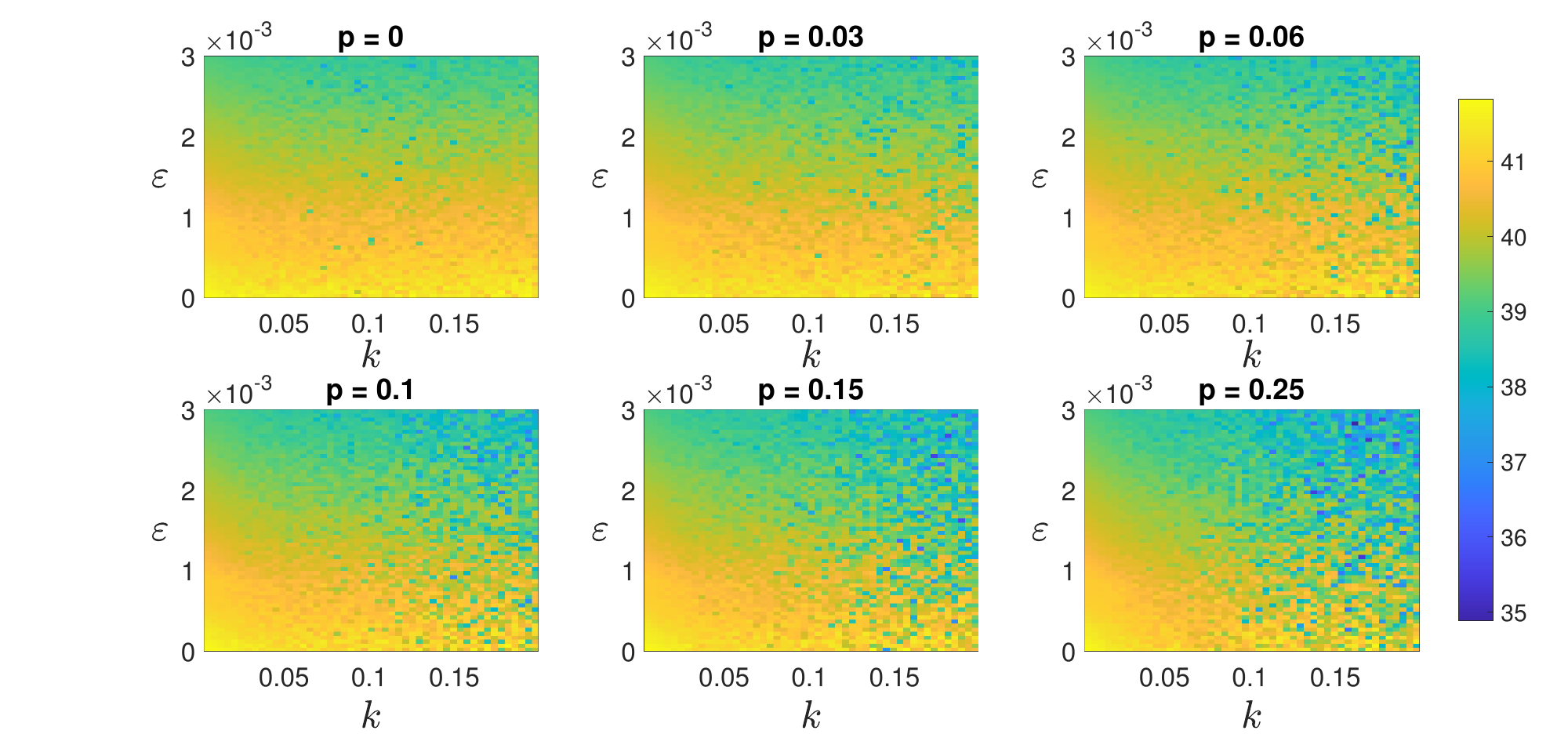}
\caption{\textbf{Plot of the inter-spike interval ($ISI$)}. Representation of the mean value of the $ISI$ for a network where 25 out of the 50 neurons have a random value for the parameter $b$. The mean value of the parameter $b=0.35$ and the mismatch is randomly selected with maximum value $\Delta b=0.01b$. The rest of the parameters are equal for all the neurons, $a=0.89$, $c=0.28$ and $I=0.03$. In the network a $5\%$ of the connections are inhibitory couplings.}
\label{fig_ISI_b_035_n2_25_inhi_05}
\end{figure}

\begin{figure}[ht!]
\centering
 \includegraphics[width=\textwidth]{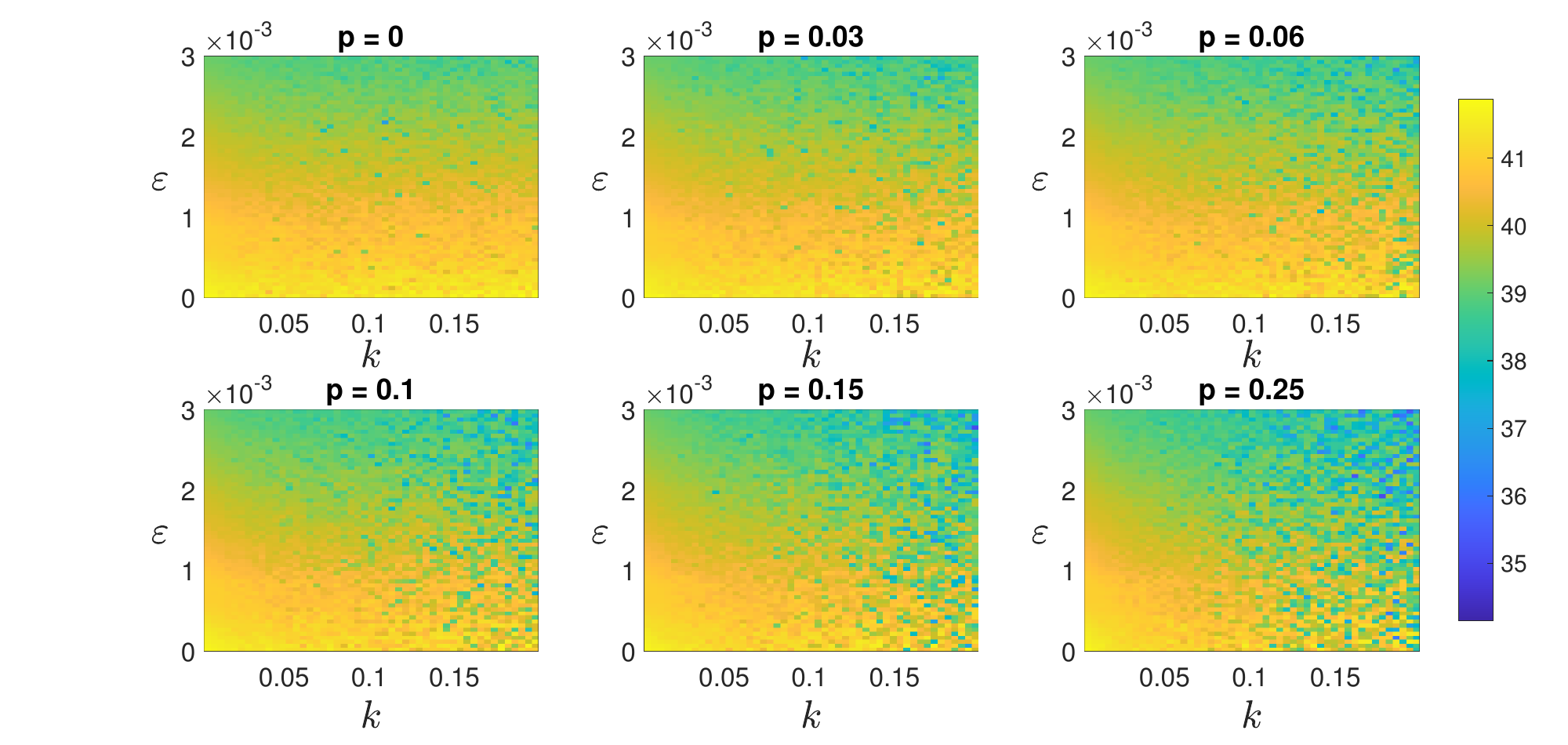}
\caption{\textbf{Plot of the inter-spike interval ($ISI$)}. Representation of the mean value of the $ISI$ for a network where all the 50 neurons from the network have a random value for the parameter $b$. The mean value of the parameter $b=0.35$ and the mismatch is randomly selected with maximum value $\Delta b=0.01b$. The rest of the parameters are equal for all the neurons, $a=0.89$, $c=0.28$ and $I=0.03$. In the network a $5\%$ of the connections are inhibitory couplings.}
\label{fig_ISI_b_035_n2_50_inhi_05}
\end{figure}

Lastly, the study of the $ISI$ is depicted in Figs.~\ref{fig_ISI_b_035_n2_0_inhi_05} to \ref{fig_ISI_b_035_n2_50_inhi_05}. Unlike the cases where all connections were excitatory or where only $1\%$ of the connections were inhibitory, the stability of the $ISI$ is worse in this scenario. Here, the range of $ISI$ values is noticeably wider.

\section{Conclusions}
 \label{conclusions}

We have numerically investigated the global behavior of a neuron network organized in the well-known small-world pattern. We used a stochastic version of the map-based Chialvo neuron model, introducing a random mismatch in the parameter $b$ in some neurons to induce heterogeneity within the network. Under these conditions, we numerically analyzed the synchronization of the network and the inter-spike interval ($ISI$) as a function of the noise intensity $\varepsilon$ introduced in the stochastic model, the coupling strength between neurons $k$, and the mismatch in the parameter $b$.
Additionally, we examined the network's global behavior based on its structure by increasing the rewiring probability of network connections $p$ and adjusting the proportion between excitatory and inhibitory connections.
In general, synchronization is more pronounced in homogeneous networks compared to heterogeneous ones. As the number of neurons with parameter mismatches increases, the system becomes more desynchronized. When the network consists solely of excitatory connections, synchronization is high and further increases with coupling strength. However, transitioning some connections to inhibitory decreases the region of high synchronization in the parameter plane ($k$,$\varepsilon$) as the proportion of inhibitory connections rises.

The most significant findings relate to both the probability of rewiring and the intensity of noise, with the latter being particularly striking, as it suggests the existence of noise levels that substantially enhance system synchronization. Regarding rewiring probability, we observe that increasing $p$ improves network synchronization. This effect remains consistent across all conditions, regardless of the proportion of neurons with mismatches or the specific types of connections within the network. Thus, we deduce that introducing randomness into the network enhances its stability.
Regarding the impact of noise on network synchronization, we identify specific noise values that lead to a marked increase in synchronization. These values are approximately centered at $\varepsilon=0.8\cdot 10^{-3}$ and $\varepsilon=2\cdot 10^{-3}$ and remain nearly constant regardless of coupling strength or the proportion of excitatory/inhibitory connections. When all connections are excitatory, the improvement in network synchronization at these noise levels is particularly evident, especially at lower rewiring probabilities. As the rewiring probability increases, the effect of noise diminishes, as the system reaches high synchronization levels. The influence of these noise levels becomes more pronounced when inhibitory connections are introduced. In both cases, optimal synchronization occurs at the aforementioned noise levels. This effect is noticeable with $1\%$ inhibitory connections but significantly more pronounced with $5\%$ inhibitory connections.

The inter-spike interval ($ISI$) of neurons remains relatively constant across all figures, regardless of coupling strength. We noted a decrease in $ISI$ as noise intensity increases, which is consistent for cases with all excitatory connections and a lower percentage of inhibitory connections (e.g., $1 \%$). However, when the proportion of inhibitory connections increases (e.g., $5 \%$), this trend is slightly modified and becomes more noticeable as the rewiring probability of network connections increases.

Finally, we observed that $ISI$ values become less homogeneous in cases with inhibitory connections and high rewiring probability. This phenomenon may be attributed to the random selection of inhibitory connections from the network, leading to internal structures that significantly affect the $ISI$ of neurons. These findings provide valuable insights for future studies investigating network synchronization and $ISI$ as functions of the network structure.

\noindent{\bf{Acknowledgments}}
JU, JMS and MAFS acknowledge financial support from the Spanish State Research Agency (AEI) and the European Regional Development Fund (ERDF, EU) under Project Nos.~PID2019-105554GB-I00 and ~PID2023-148160NB-I00 (MCIN/AEI/10.13039/501100011033). IB and LR acknowledge financial support from the Ministry of Science and Higher Education of the Russian Federation (project ‘‘Ural Mathematical Center’’, Agreement N 075-02-2025-1719/1).

\end{document}